\documentclass[sigconf]{acmart}
\AtBeginDocument{%
  \providecommand\BibTeX{{%
    \normalfont B\kern-0.5em{\scshape i\kern-0.25em b}\kern-0.8em\TeX}}}

\copyrightyear{2025}
\acmYear{2025}
\setcopyright{rightsretained}
\acmConference[CHI '25]{CHI Conference on Human Factors in Computing Systems}{April 26-May 1, 2025}{Yokohama, Japan}
\acmBooktitle{CHI Conference on Human Factors in Computing Systems (CHI '25), April 26-May 1, 2025, Yokohama, Japan}\acmDOI{10.1145/3706598.3713969}
\acmISBN{979-8-4007-1394-1/25/04}

\newcommand{\rv}[1]{\textcolor[HTML]{000000}{#1}}

\newcommand{\rvtwo}[1]{\textcolor[HTML]{000000}{#1}}

\usepackage{ifpdf}%
\ifpdf
  \usepackage{epstopdf}%
\fi
\usepackage{colortbl}
\usepackage{subcaption}
\usepackage{multirow}
\usepackage{makecell}
\usepackage{wrapfig}
\usepackage{enumitem}
\usepackage{url}
\usepackage{hanging}
\usepackage{stfloats} 
\setlist[itemize]{noitemsep, topsep=0pt}
\begin{document}


\title[Leveling Up Together: Positive and Safe Space for Teen Developers]{Leveling Up Together: Fostering Positive Growth and Safe Online Spaces for Teen Roblox Developers}

\author{Yubin Choi}
\email{joyda2525@gmail.com}
\orcid{0000-0002-6042-5313}
\affiliation{%
  \institution{KAIST}
  \city{Daejeon}
  \country{Republic of Korea}
}

\author{Jeanne Choi}
\email{jeannechoi@kaist.ac.kr}
\orcid{0009-0005-7183-199X}
\affiliation{
  \institution{KAIST}
  \city{Daejeon}
  \country{Republic of Korea}
}

\author{Joseph Seering}
\email{seering@kaist.ac.kr}
\orcid{0000-0001-7606-4711}
\affiliation{
  \institution{KAIST}
  \city{Daejeon}
  \country{Republic of Korea}
}

\renewcommand{\shortauthors}{Choi and Choi, et al.}

\begin{abstract}
    Creating games together is both a playful and effective way to develop skills in computational thinking, collaboration, and more. However, game development can be challenging for younger developers who lack formal training. While teenage developers frequently turn to online communities for peer support, their experiences may vary. To better understand the benefits and challenges teens face within online developer communities, we conducted interviews with 18 teenagers who created games or elements in Roblox and received peer support from one or more online Roblox developer communities. Our findings show that developer communities provide teens with valuable resources for technical, social, and career growth. However, teenagers also struggle with inter-user conflicts and a lack of community structure, leading to difficulties in handling complex issues that may arise, such as financial scams. Based on these insights, we propose takeaways for creating positive and safe online spaces for teenage game creators. 
\end{abstract}




\received{September 2024}
\received[revised]{December 2024}
\received[accepted]{January 2025}

\maketitle

    %\captionsetup{belowskip=0pt, aboveskip=0pt}
\section{Introduction}

\begin{figure}[!htbp]
    \centering
    \includegraphics[width=0.8\linewidth]{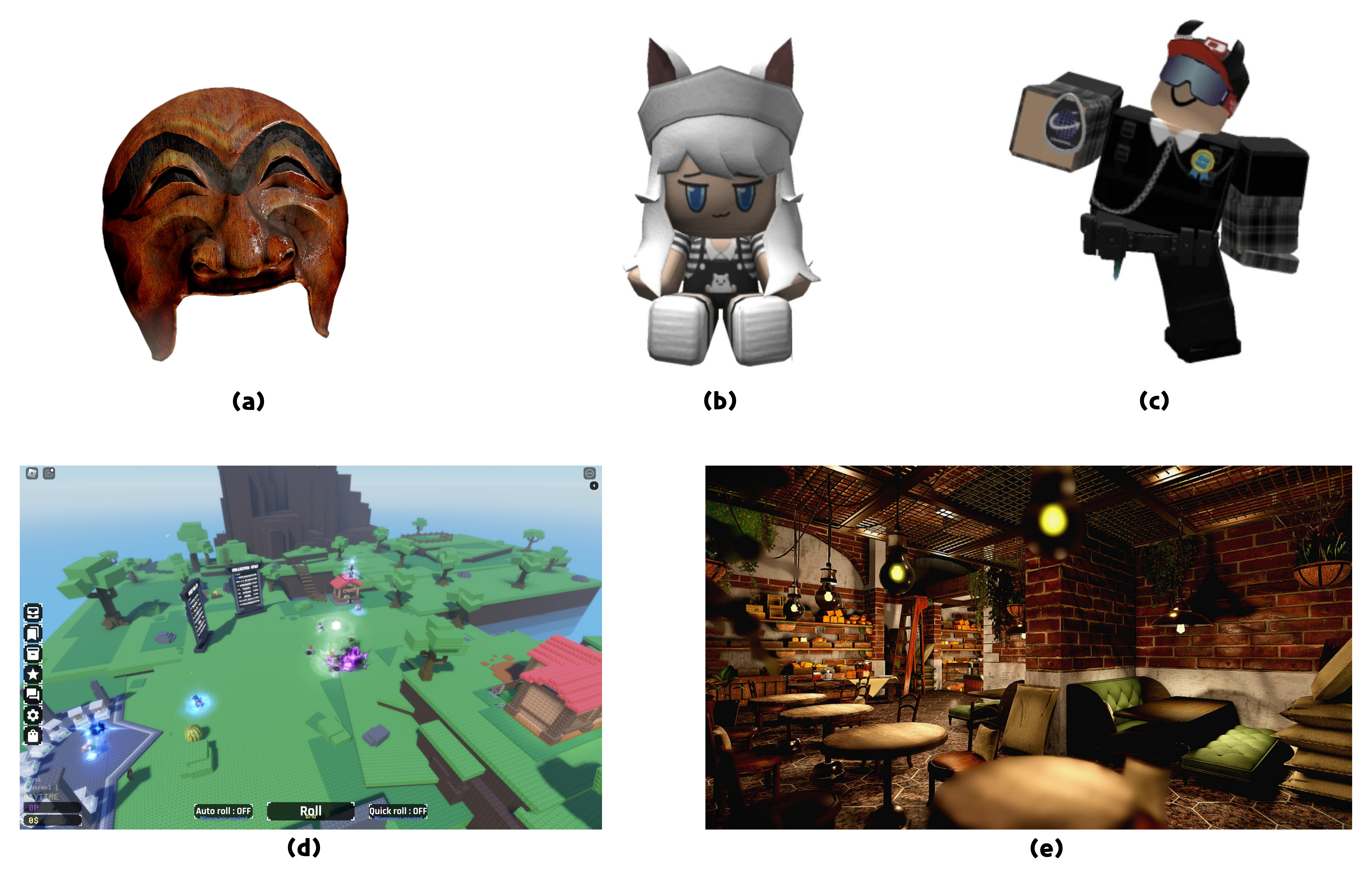}
    \caption{Games and elements made by study participants shared upon their permission. (a) and (b) are User Generated Content (UGC) that could be added to decorate personal avatars like (c). The images of (d) and (e) are game spaces.}
    \label{fig:creation}
    \Description{Five images labeled (a), (b), (c), (d), and (e) show the Roblox creations of the participants of the user study, which are User Generated Content (UGC). Images (a), (b), and (c) are 3D items, where (a) is a traditional Korean wood mask, (b) is a woman 3D character with a cat-ear hat sitting down, and (c) is a main character with sunglasses walking. Images (d) and (e) both are screenshots of a 3D Roblox space, where (d) is during the game in a grass field and (e) is a vintage interior space.}
\end{figure}

\rv{Game development} offers teenagers an engaging and valuable opportunity to build technical skills, practice collaboration, and develop self-identity~\cite{davison2006adolescent, roseth2008promoting}. \rv{Advances} in programming support tools and school coding education have lowered the barrier \rv{for adolescents to start creating their own games.} \rv{Yet, game development} still \rv{remains} a \rv{complex endeavor which requires a diverse set of skills, including design and programming skills as well as communication skills to engage with player feedback after deployment}. Online developer communities \rv{have emerged as an important environment for supporting these young creators, providing spaces to} share knowledge, \rv{overcome technical hurdles, and} learn from peers~\cite{10.1145/2531602.2531659, 8417152, brennan2021kids}. \rv{These communities can break down geographic and hierarchical boundaries, fostering connections through shared expertise~\cite{hwang2015knowledge}}. However, online developer communities are often \rv{created by and} run for \rv{adult game developers, which may cause them to be less suitable for teen developers, potentially leading to risks of exposure to content and behaviors these young developers are not yet ready to handle.} Prior \rv{research highlights how teenagers are acutely vulnerable to the potential harms they may face in navigating online communities ranging from bullying to harassment and exposure to extreme content~\cite{harlan2012teen, subrahmanyam2008online, 10.1145/2531602.2531659, heung2024vulnerable, park2022measuring, kumar2023understanding, chatterji2016code}.}

To \rv{better} understand this dual nature of online developer communities for teenager games creators---both their potential value and the risks they pose---we examined the case of \rv{Roblox} developer communities. \rv{As a} global game platform, \rv{Roblox} hosts more than 40 million games\footnote{\url{https://backlinko.com/roblox-users}} and holds a significant \rv{role in} youth culture, with 58\% of its 79.5 million daily active users under the \rv{age of 16}.\footnote{\url{https://create.roblox.com/creator}} \rv{Compared to} previously studied environments such as the Minecraft modding ecosystem ~\cite{10.1145/2858036.2858038, grace2014towards, slovak2018mediating} and the Scratch programming learning-focused platform~\cite{10.1145/3491102.3502124, brennan2021kids, shorey2021hanging, roque2016children}, \rv{Roblox stands as a digital environment where teenagers have unparalleled autonomy to create, explore, and monetize their content with the following reasons.} \rv{First,} Roblox \rv{supports a diverse range of creators with different levels of experience}. Content on Roblox, from full-fledged games (known as “Experiences”) to individual character assets like avatar clothing and accessories \rv{(known as “Creations”) and game assets like background music, is predominantly user-generated. Therefore, we define ``Roblox game developers'' to broadly include not only game developers in the traditional sense but also designers, asset creators, and special effect artists.}  \rv{Second, }Roblox \rv{supports developers} through ``Roblox Studio'', \rv{a beginner-friendly platform for}  generating and publishing their user-generated content with \rv{Lua scripting, a programming language known for its simplicity.} \rv{Third, Roblox's monetization model allows developers to turn their creativity into tangible rewards. Their }in-platform currency, “Robux,” can be converted into real-world money based on their specific policies.\footnote{Robux can be exchanged for
%real-world currency under certain conditions.}
\rv{0.0013 USD per 1 Robux as of the time of writing, after certain thresholds have been met.}}
\rv{Through this autonomy and accessibility, teen developers can engage in playful experimentation, gain independence, and build peer connections. This redefines the dynamics of user-generated content creation, providing a compelling case to examine both opportunities and challenges for teen developers.} 

\rv{To support its developers}, Roblox \rv{maintains} an official online community, the “DevForum,” \rv{and official social media channels in platforms like Discord. However, individual developers have also established numerous unofficial, user-driven communities.} These spaces cater to various developer needs, fostering peer-to-peer learning and collaboration. By studying Roblox’s developer communities, we aim to \rv{understand both} how online spaces can empower teenage creators \rv{and also how they can potentially expose} them to risks. \rv{Our research was} motivated \rv{by the unique context of Roblox developer communities, where teenage developers} exercise significant agency and autonomy. \rv{Therefore}, our research focuses on three core questions:

%To investigate the experiences of teenage developers within these online communities

\begin{itemize}
    \item \textbf{RQ1. Community Use:} \rv{Why} do teen Roblox developers utilize online developer communities? 
    \item \textbf{RQ2. Benefits:} What benefits do \rv{teen Roblox developers} receive from participating in these communities and how do the benefits vary across different creators?
    \item \textbf{RQ3. Challenges:} What challenges do teen Roblox developers face within these communities and \rv{what strategies do they use to cope} with these challenges?
\end{itemize}

We conducted semi-structured interviews with 18 teen Roblox developers to \rv{explore their experiences in their own words}. Our findings reveal that these communities \rv{function} as a dynamic digital playground where \rv{teen} developers explore ideas, collaborate with peers, and \rv{may even find the beginnings of a career in game development by leveraging the monetization opportunities that Roblox provides}. However, these communities also present considerable challenges. The \rv{interviewed teen} developers reported encountering financial scams, inappropriate user \rv{interactions}, difficulties balancing their online activities with their schooling, and being discouraged by the \rv{stigma surrounding the} child-like image of Roblox. By examining these experiences, our study aims to provide insights into how online communities can better support \rv{teenaged developers}, fostering a productive environment that maximizes their creative potential while \rv{proactively addressing} the associated \rv{safety risks}.

\section{Related Work}
Our paper builds on three primary bodies of prior work. First, we explore research on game development with young users.  Second, we discuss research on the social dynamics of online developer communities. Finally, we present prior work on risks and challenges for youth \rv{creators} in online communities.

% Sign posts. 1) Remind readers - what to expect beginning , 2) heres why I'm telling u this at the end
% Write as story setting = related work

\subsection{Youth Game Design and Development as Digital Play}
Play is widely recognized as a crucial process for children's learning and development~\cite{singer2006roberta}. In today's digital age, the concept of play extends from physical to virtual environments like TikTok, YouTube, and various social media platforms~\cite{10.1145/3173574.3174233}. \rv{Resonating with theories of intrinsic motivation and participatory culture~\cite{jenkins2009confronting}, an increasing number of teenagers have become both consumers and producers--termed ``prosumers'' in digital spaces, participating by creating their own content in the form of blogs, gameplay streams, and game mods~\cite{harlan2012teen, lombana2020youth}.} While these ``digital playgrounds'' provide new avenues for engagement \rv{and learning for youth}, they are mainly designed by adults to meet adult-centric goals and constraints. However, the digital autonomy of children and adolescents --- the freedom to explore and express themselves in digital spaces---can be important, enabling children to gain independence, build peer relations, and form a self-identity~\cite{davison2006adolescent, roseth2008promoting, wang202312}. Game development by the youth provides a unique opportunity to achieve this autonomy by creating and exploring their own games or virtual worlds, offering a type of playful experimentation not available on traditional adult-focused platforms.

Extensive research in HCI has shown that children and teenagers can benefit from learning to design and create games. Recent advances in technology~\cite{10.1145/3544548.3580976, 10.1145/2470654.2481360, 10.1145/3613904.3642895}, combined with the rise of computational education in schools~\cite{rijo2022computational, bers2022state}, and the development of child-friendly programming tools like ScratchJR, and Lego Mindstorms~\cite{flannery2013designing, kert2020effect}, have lowered the barriers to game design. This democratization of game creation, driven by indie developers and modding~\cite{10.1145/3375184, freeman2019exploring, sotamaa2010game} is complemented by studies showing children can also make games. Studies have demonstrated that involving children as co-designers in ``serious'' games helps them better grasp complex real-world issues~\cite{tucker2019broke}. Further, children learn computational concepts~\cite{10.1145/3544548.3581272, 10.1145/3025453.3025847, 10.1145/3311927.3323129} through both ``serious''~\cite{10.1145/3613905.3650833, kert2020effect} and ``non-serious''~\cite{troiano2019my} game design. Collaborative game design has also been linked to improved communication skills and higher \rv{learning involvement} of children~\cite{denner2014pair, gritschacher2012standing}. %Also, a massive collaboration between children, on a LEGO-style website, was highly motivated by entertaining their friends and learning from other children~\cite{gritschacher2012standing}. 
%Further, children who were self-motivated in interest to learn concepts such as data structures in Scratch had better learning outcomes~\cite{10.1145/3491102.3502124}. 
Broadly, giving teenagers autonomy in creating games leads to learning benefits across a variety of domains~\cite{10.1145/3491102.3502124, gritschacher2012standing}. We focus on Roblox, for its flexibility in game design options, kid-friendly programming language (Luau), and tools (Roblox Studio), making it a strong platform for teen-led game development. In this study, we explore the potential benefits that teen game designers \& developers can gain from participating in the supportive social ecosystem of creating games.

\subsection{Online Developer communities}
%1. Developer communities are valuable spaces for support-seeking. 2. A lot of research has studied developer spaces focused around adults, such as Github teams, and found benefits X, Y, and Z.

Online developer communities \rv{built around platforms such as Github provide essential} spaces for learning, \rv{collaboration}, and social interaction. Development \rv{is rarely a straightforward process, and these communities often become hubs for mutual support.} Much research on adult-focused developer communities, such as GitHub and StackOverflow, has found that these communities \rv{provide} valuable opportunities for knowledge sharing~\cite{10.1145/2531602.2531659, 8417152}, enhancing individual expertise~\cite{may2019gender}, and collaborative coding to complete projects~\cite{vasilescu2014continuous}. \rv{Similarly, indie game developer communities serve as \textit{communities of practice}, where small, flexible team dynamics foster autonomy and creativity. In these settings, members often adapt their roles democratically, emphasizing collaboration and shared ownership~\cite{freeman2019exploring}.}

\rv{While much of the prior work in this space focuses on \textit{adult} developer communities, research on teen-focused developers, particularly in the context of Scratch, has uncovered a breadth of opportunities and challenges.} Studies reveal the benefits of developer communities in learning computational skills and exchanging knowledge with peers~\cite{brennan2021kids}. \rv{Further, in Scratch team challenges, studies demonstrate that youth deploy sophisticated socialization skills: negotiating dynamic leadership styles for each project, building interpersonal trust, and cultivating common ground~\cite{fields2013understanding, aragon2009tale, chou2018designing}.} However, \rv{youth face distinct communication obstacles compared to adults. They often struggle to} articulate their thoughts~\cite{10.1145/3491102.3502124, roque2016children} and adapt to the \rv{nuanced} engagement \rv{expectations among different online communities}~\cite{fields2015have}. \rv{Critically, peer recognition and feedback are pivotal motivations for community engagement—universally important, but especially crucial for teenaged developers}~\cite{kafai2012collaborative}. \rv{Young creators, still developing their creative identities and networks, rely more intensely on peer recognition and feedback from engagement. These social interactions are fundamental to driving participation, building confidence, and fostering belonging within digital learning environments.}
%Recognition and feedback from peers play critical roles in these communities, as they motivate participation and support members’ sense of belonging~

\rv{In contrast to previously studied Scratch communities, Roblox developer communities, while similarly youth-driven, operate within a distinct social and economic framework. Unlike Scratch, which prioritizes education and creativity, Roblox combines social engagement with broader agency in creations and a deeply embedded in-platform monetization ecosystem. Work by Zhang et al. has shown that Roblox teen developer communities engage in forms of collective ideation, discussing game design strategies and sharing feedback through external social media platforms such as Reddit~\cite{zhang2024understanding}. Taken together, these findings underscore the importance of understanding online developer communities as dynamic environments that not only facilitate technical skill development but also nurture collaboration and social belonging for teen developers.}
\rv{While, previous research} has primarily analyzed teen developers' \rv{public comments and forum interactions, our study directly engages teen developers. By conducting first-hand interviews, we uncover rich, nuanced insights into their lived experiences, motivations, and the challenges of participating within these youth-driven developer communities}.

%However, prior has focused primarily on external discussions and has not captured the firsthand experiences of teen developers across different communities. By directly engaging with youth through interviews, our study provides deeper insights into their experiences, enabling us to understand how they perceive the benefits and challenges of participating in these youth-driven spaces.}

%This paper examines Roblox developer communities, which have significant appeal to younger audiences voluntarily more focused on engagement than education in Scratch and thus may have significantly different social dynamics than developer communities with primarily older users. In contrast to previously studied youth-led programming spaces like Scratch, Roblox features active in-platform monetization, which complicates the motivations for participation.

%In addition, adult-oriented developer communities may not have the type of patience for the processes of identify formation that youth-oriented communities may be designed to support~\cite{ringland2016will}.

\subsection{Child safety issues in online developer communities}
%J: 1. While there are many benefits from participation in online developer communities, there are also challenges and risks surrounding conflict and safety. 2. Previous work has identified challenges/safety risks A, B, and C in online developer communities.

Participation in online developer communities offers benefits, but \rv{it may also expose developers to} a variety of challenges and risks. Research has \rv{consistently} highlighted issues \rv{in online social spaces} such as toxic comments, name-calling, and offensive language~\cite{10.1145/3510003.3510111, ferreira2021shut, cheriyan2021towards}, as well as programming-specific risks like code theft~\cite{chatterji2016code} and malicious hacking in shared code~\cite{kera2012hackerspaces}. Another layer of risks \rv{emerges} in developer communities on social media platforms like Reddit and Discord, \rv{where many Roblox developer communities are based}. These platforms bring \rv{the} traditional threats \rv{that have been identified in studies of social media moderation}, including discrimination~\cite{heung2024vulnerable} and personal attacks~\cite{park2022measuring} \rv{to developers}. For example, members of Reddit communities may experience repeated harassment across multiple subcommunities~\cite{kumar2023understanding}. Children are particularly vulnerable \rv{in these environments due to the possibility of encountering more severe risks such as} sexual abuse, cyber grooming, and exposure to explicit content~\cite{ali2021child, freed2023understanding}. \rv{Exposure to} explicit material \rv{within developer communities} can lead to \rv{long-term detrimental effects on mental health}~\cite{tanni2024examining}. Furthermore, \rv{prior work has suggested that adolescents may face emotional challenges stemming from addiction to or dependence on these platforms}~\cite{wisniewski2015resilience}.

\rv{The Roblox ecosystem presents unique challenges due to its distinctive community structure. Unlike traditional platforms, Roblox blurs the boundaries between players and developers, creating a complex social dynamic where young creators often emerge directly from the player community. Kou et al. have pointed out that policies being unclear between creators and players as one of the reasons for harmful games including those featuring racist or misogynist content\cite{kou2023harmful}. This fluid transition can inadvertently normalize problematic content, as creators may unconsciously draw inspiration from games that incorporate harmful themes or design elements. Also, the platform's proprietary development environment creates a form of technological "lock-in" that constrains monetization options only within Roblox. This can pressure teen developers to prioritize engagement and monetization over ethical design considerations. Moreover, what researchers term "aspirational labor" drives young creators to invest significant time and effort into content creation, often with minimal compensation~\cite{lombana2020youth}. The promise of potential future success motivates children to engage in unpaid creative work. Media critiques have pointed out Roblox's potentially exploitative practices of relying on young users to create content without proper compensation~\cite{The_Guardian_2022}. }

\rv{Though there has been much prior literature on online safety and aspirational labor, limited research has explored the overlapping dynamics of these phenomena in platforms like Roblox, where player and developer roles merge in communities where youth have significant agency. Our study addresses this gap by examining how youth perceived and experienced potential risks faced by teen developers, highlighting the major risks and proposing guidelines for potential interventions that balance opportunities and safeguards in youth-driven digital economies.}

%Moreover, research has identified several safety issues within the Roblox environment, including exposure to viruses, predatory behavior, harmful game designs, and exploitation. For instance, Kou et al., through analysis of Reddit comments, found that developers faced malicious activities and that the proprietary nature of Roblox’s system presents challenges, as the ecology of its unique language and tools created a form of lock-in.~\cite{kou2024ecology}. Also, Kou's study on Reddit found that harmful game designs existed with inadequate moderation of problematic content like racism or misogyny~\cite{kou2023harmful}. Furthermore, media have criticized Roblox for exploitative practices of relying on young users to create content without proper compensation~\cite{The_Guardian_2022}.

%Together, these findings underscore the multifaceted risks teenage Roblox developers face in communities and highlight the need for deeper exploration of how teen developers are dealing with the risks and ways to safeguard the well-being of young developers.

\section{Methods}
Our goal for this paper was to understand the benefits and challenges teenager Roblox creators encountered in online communities. To this end, we interviewed participants who have joined at least one online developer community and have self-identified as "creators" in Roblox.

\subsection{Participants}
We recruited participants who met three primary criteria: 1) Teenagers between the ages of 13 - 19,\footnote{\rv{Participants in South Korea were recruited up to age 19, as 19 is the age of majority in South Korea according to legal and cultural norms. Non-Korean participants were recruited up to age 18.}} 2) ``Creators'' who have created two games or three game elements, per Roblox's definition of having created development tools (games, maps, models, or assets) or avatar items (User Generated Content, UGC), 3) being a member in at least one online developer community related to Roblox. \rv{We included participants aged 13–19 to capture a broad spectrum of experiences, as online behaviors and risks vary across different stages of adolescence~\cite{diaz2024dysfunctional}.} To verify that participants were creators, we required them to share their creations with us during the interview.
%as part of the recruitment process. 
%\textcolor{white}{participated} 

To recruit such participants, we posted interview calls in popular Roblox online communities. We first joined online communities that showed up in a Google search for Roblox developer communities, including the Roblox official Developer Forum (DevForum), three Discord servers, and one Reddit subcommunity (r/robloxgamedev). Within those communities, we posted interview calls after getting permission from the moderators. We also recruited participants via snowball sampling, and we posted new calls in communities we learned about in early interviews. In total, calls were posted on the DevForum English and Korean Recruitment pages, nine Discord servers, one Reddit subcommunity, and one Naver Cafe --- a social platform popular in Korea. Participants who were willing to be interviewed filled out a pre-survey that was designed to check whether participants matched our recruitment criteria. As a result, 21 participants were recruited. One participant dropped as she did not want to share what she made and two did not fit our criteria. In the end, we interviewed 18 participants. 
In Table ~\ref{tab:demographics}, we show the demographics and information about our participants. The mean age of the participants was 16.3 years old (SD = 1.3) with a range of 14--19 years old. All participants were high school students except C15 who was enrolled in college at the time of the interview. Sixteen participants identified as male, and two participants identified as female. While there is no official census on the developer gender distribution of Roblox, statistics from StackOverflow developers show that female participants may be significantly underrepresented in developer communities. Of the 18 participants, 15 identified themselves as skilled in Scripting (programming) and 7 in 3D Modeling (multiple-choice possible). The average years of experience as a Roblox creator was 3.4 years (details \rv{in} Table ~\ref{tab:demographics}). 11 Koreans, 3 Americans, and one \rv{participant from each of Malaysia, Spain, Japan, and Denmark} were recruited. While we had originally intended to report how long each participant had been in the developer community they had joined, we found that most participants had joined multiple communities (with one participant reporting having joined over 200) and could not recall the exact dates of joining. Therefore, instead, we report the platforms that each participant joined communities through.

\subsection{Semi-structured Interview Method}
We conducted semi-structured interviews to understand how teenager Roblox creators used online developer communities, the benefits they gained, and the challenges they faced. Interviews generally proceeded in four phases: 1) warm-up questions, 2) experiences as a creator, 3) developer community usage, benefits, and challenges, and 4) offline impact of Roblox developer experience. Example questions include \textit{"Where did you go when you were stuck when creating {previously mentioned creation}?", "What would you miss if {joined online community} did not exist?" "Have you shown what you made to your parents?"}. \rv{The full interview protocol is shown in Appendix \ref{appendix:questions}}. All interviews were conducted online via Discord Call or Zoom (without video) from the 3rd of July to August 15th, 2024 by the first author. The interviews were conducted in either English or Korean based on the interviewee's preference. The average length of the interview was 77 minutes. All participants received \$20 (or equivalent in local currency) for participating in the interview. 

%longer than the scheduled 60-minute block, as participants were very enthusiastic to share their experiences. 

%\input{tables/demographics}

\begin{table*}[ht]
\resizebox{.6\textwidth}{!}{
\begin{tabular}{ccc}
\Xhline{1pt}
\rowcolor[HTML]{EFEFEF} 
\textbf{Demographics}                                                                                                                          & \textbf{Options}                                               & \textbf{Count }(n=18)                                                                                           \\ \Xhline{1pt}
\textbf{\begin{tabular}[c]{@{}c@{}}Age\end{tabular}}                                                                                 & Number                                                         & \begin{tabular}[c]{@{}c@{}} $Mean$ = 16.3 years\\ $SD$ = 1.3 years \\ $Range$ = 14-19 years\end{tabular}                                           \\ \hline
\rowcolor[HTML]{F3F3F3} 
\cellcolor[HTML]{F3F3F3}                                                                                                                       & Male                                                           & 16                                                                                                       \\
\rowcolor[HTML]{F3F3F3} 
\multirow{-2}{*}{\cellcolor[HTML]{F3F3F3}\textbf{Gender}}                                                                                      & Female                                                         & 2                                                                                                        \\ \hline
                                                                                                                                               & South Korea                                                    & 11                                                                                                       \\
                                                                                                                                               & United States                                                  & 3                                                                                                        \\
                                                                                                                                               & Malaysia                                                       & 1                                                                                                        \\
                                                                                                                                               & Japan                                                          & 1                                                                                                        \\
                                                                                                                                               & Spain                                                          & 1                                                                                                        \\
\multirow{-6}{*}{\textbf{Nationality}}                                                                                                         & Denmark                                                        & 1                                                                                                        \\ \hline
\rowcolor[HTML]{F3F3F3} 
\cellcolor[HTML]{F3F3F3}                                                                                                                       & Scripting                                                      & 15                                                                                                       \\
\rowcolor[HTML]{F3F3F3} 
\cellcolor[HTML]{F3F3F3}                                                                                                                       & 3D Modelling                                                   & 7                                                                                                        \\
\rowcolor[HTML]{F3F3F3} 
\cellcolor[HTML]{F3F3F3}                                                                                                                       & Building                                                       & 5                                                                                                        \\
\rowcolor[HTML]{F3F3F3} 
\cellcolor[HTML]{F3F3F3}                                                                                                                       & GFX                                                            & 2                                                                                                        \\
\rowcolor[HTML]{F3F3F3} 
\cellcolor[HTML]{F3F3F3}                                                                                                                       & Animation                                                      & 1                                                                                                        \\
\rowcolor[HTML]{F3F3F3} 
\multirow{-6}{*}{\cellcolor[HTML]{F3F3F3}\begin{tabular}[c]{@{}c@{}}\textbf{Profession}\\ \textit{(multi-choice possible)}\end{tabular}}            & Game Design                                                    & 1                                                                                                        \\ \hline
\textbf{\begin{tabular}[c]{@{}c@{}}Roblox Creation \\ Experience\end{tabular}}                                                                 & Number                                                         & \begin{tabular}[c]{@{}c@{}}$Mean$ = 3.4 years\\ $SD$ = 2 years\\ $Range$ = 4 months $\sim$7 years\end{tabular} \\ \hline
\rowcolor[HTML]{F3F3F3} 
\cellcolor[HTML]{F3F3F3}                                                                                                                       & Discord                                                        & 17                                                                                                       \\
\rowcolor[HTML]{F3F3F3} 
\cellcolor[HTML]{F3F3F3}                                                                                                                       & DevForum  & 12                                                                                                       \\
\rowcolor[HTML]{F3F3F3} 
\cellcolor[HTML]{F3F3F3}                                                                                                                       & Naver Cafe                                                     & 4                                                                                                        \\
\rowcolor[HTML]{F3F3F3} 
\cellcolor[HTML]{F3F3F3}                                                                                                                       & KakaoTalk                                                          & 7                                                                                                        \\
\rowcolor[HTML]{F3F3F3} 
\cellcolor[HTML]{F3F3F3}
\multirow{-4}{*}{\cellcolor[HTML]{F3F3F3}\begin{tabular}[c]{@{}c@{}}\textbf{Participating Communities}\\ \textit{(multi-choice possible)} \end{tabular}} & Reddit                                                         & 1                                                                                                        \\ \Xhline{1pt}
\end{tabular}
}
\caption{Participant information overview. To anonymize participants, we present aggregate statistical data to describe them and refer to each person with an alias C1-C18. DevForum is managed by Roblox Corp. Other communities are managed by individuals or teams of volunteers.}%Alternative for anonymizing participants more. }
\label{tab:demographics}
\Description{A table that shows the information of the user study participants. The first column is the list of demographics, which are age (14 to 19), gender, nationality, profession in Roblox development, years of Roblox creation experience, and the communities that participants use. The major data of the table is as follows. The mean age is 16.3, the majority gender is male, the majority nationality is South Korea, the majority profession is scripting, the mean Roblox creation experience is 3.4 years, and the main communities that the participants used are Discord and DevForum.}
\end{table*}

\subsection{Ethical considerations}
This study was approved by the KAIST Institutional Review Board. However, as we involved teenage participants, we took additional precautions beyond those required by our institution. Participants were sent the interview questions in advance, and before the interview, we again explained the interview process thoroughly to ensure that they fully understood the interview contents and \rv{how} the data would be collected for payment and analysis. Also, we emphasized that participants could skip a question or end the interview if they did not want to answer. Further, for payment, all participants received their parent or guardian's signature. After receiving the participants' consent, we proceeded with the interviews.

\subsection{Analysis}
We conducted inductive thematic coding~\cite{braun2006using} \rv{following the six steps of Braun's method on interview transcripts to understand the experiences of teens in Roblox developer communities and the benefits and challenges they perceived}.
%to analyze the transcriptions. 
The process started with uploading all audio recordings of the interview to Dovetail.\footnote{\url{https://dovetail.com/}} The transcriptions were automatically done in English or Korean based on the interview language by Dovetail. Then, the first and second authors each \rv{open-}coded six interview transcripts independently while correcting inaccurate transcriptions along the way, \rv{identifying} low-level \rv{themes} \rv{focusing on how various online communities support teen developers of Roblox but also the challenges they face as a teenager}. Next, the two authors discussed the \rv{themes} to reduce the initial \rv{themes} from 587 low-level concepts to 15 higher-level themes. The final themes were merged or split iteratively via Dovetail through rounds of discussion \rv{on overlapping themes or disagreements}. 
%After agreeing on the final codes, 
\rv{Based on the refined themes,} the two interviewers \rv{engaged in another iterative discussion regarding remaining inconsistencies, disagreements, and newly emarging themes, re-coding} two interviews (C5, C12) independently again \rv{to confirm agreement}. {Cohen's Kappa was calculated to determine} inter-rater agreement, \rv{reaching a value of 0.78}. After reaching agreement \rv{on the themes}, the first author coded the remainder of the interviews. The \rv{final list of themes} can be seen in Appendix~\ref{appendix:codebook}.

\subsection{Limitations}
Before presenting our results, we acknowledge several limitations of this paper. First, our participant sample does not fully represent the global population of teen Roblox developers. Though we do not know official statistics on teen Roblox developers, there is an absence of participants from South America, Africa, and other regions \rv{in our sample (See Table ~\ref{tab:demographics})}. Since different cultures have communities operating in their languages, our findings are constrained by the language capabilities of the research team. Moreover, even within the same language, each community has unique norms; for this study, we recruited primarily from larger communities with more than 10,000+ participants, which may limit the generalizability of our findings. 

Further, because our recruitment calls included criteria that required participants to have made at least two creations, our participants are likely skewed toward those who are more active community members. As seen in Table ~\ref{tab:demographics}, the majority of the participants in this study had spent more than two years creating content for Roblox, so true newcomers who had just begun creating are underrepresented in our sample. 

\section{Results}
\rv{In this section, we summarize why teen developers joined online communities and which ones they joined. We report the benefits they found and the challenges they faced during their community experiences. For each challenge, we discuss participants' coping strategies.}

\subsection{How do teen Roblox developers utilize online developer communities? (RQ1)}
\rv{This section first introduces the motivations that drive teen Roblox developers to participate in developer communities. Next, we show how teen developers use the communities to meet multifaceted technical, social, and career needs, depending on the distinct characteristics of each community.}
\begin{figure*}
    \centering
    \includegraphics[width=0.9\linewidth]{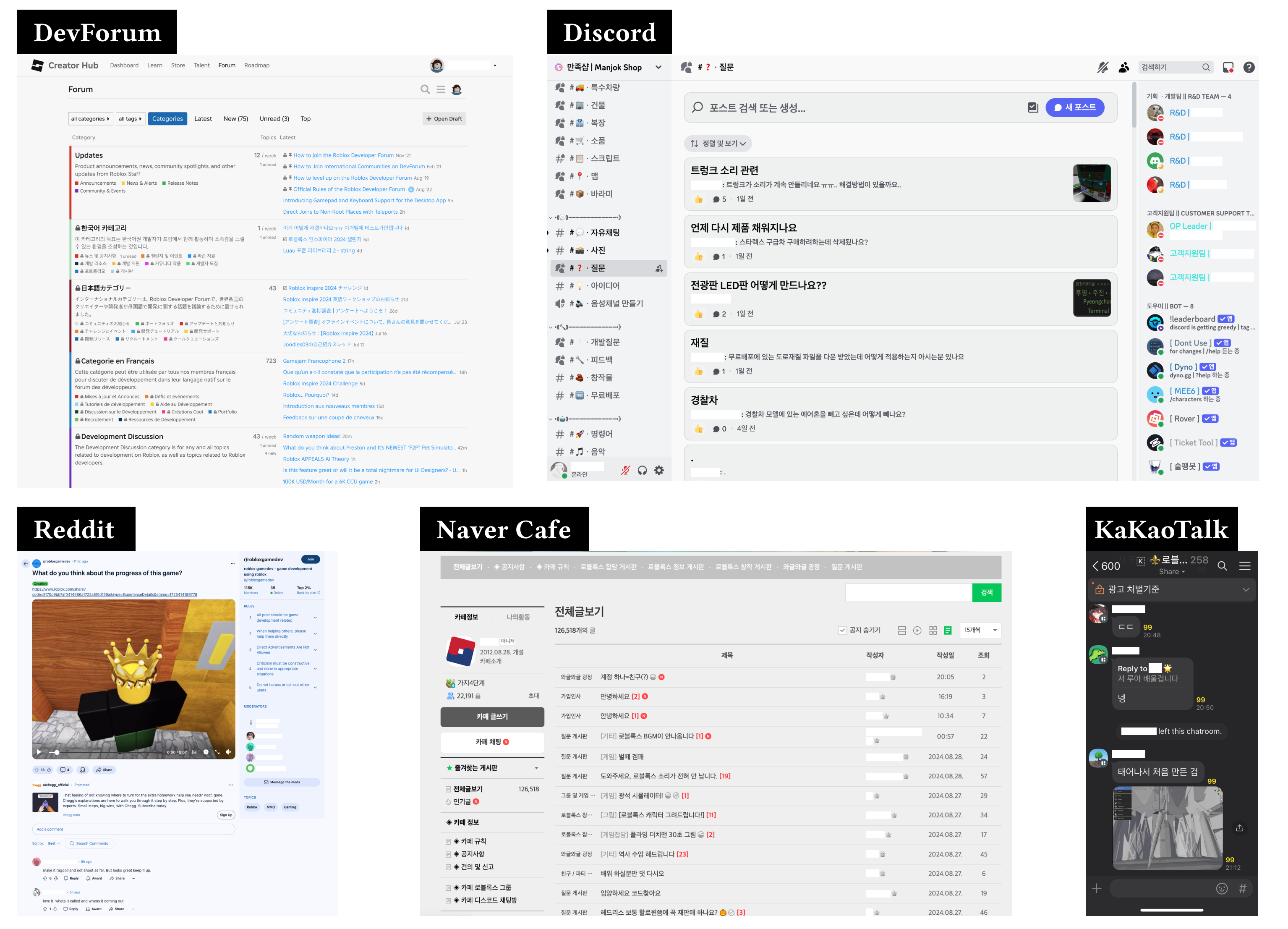}
    \caption{Different online developer communities}
    \label{fig:communities}
    \Description{Five screenshots of various online developer communities, which are DevForum, Discord Server, Reddit, Naver Cafe, and KaKaoTalk. Each screenshot shows interactions between users within the community.}
\end{figure*}

%\rv{Note: Which communities they're joining -> what triggers them to join a new community? }\\
 
\indent \rv{\textbf{Participants joined communities to get help or for social support.}} \rv{For teen Roblox developers, the journey into online communities often started with a spark of technical necessity and social curiosity. For example, C12 started with a simple search: “Roblox scripting help.” Stuck on a bug that refused to cooperate, they found themselves on DevForum, which appeared in the top 3 results in a Google search. Others, like C7, found their way into communities through a friend or searching popular social media platforms to find somebody to talk to about Roblox development. Per C2, \textit{“my school friend sent me an invite to this scripting server. At first, it was just for fun, but then I realized I could ask questions there too. It wasn’t just work—it felt like hanging out.”} Korean participants like C16 also mentioned searching open chat rooms in KaKaoTalk or Naver Cafe, which are both frequently used social media platforms in Korea, for a friend to talk about Roblox development.}\\

%DevForum, with its structured format and wealth of archived discussions, was often the first stop.}

\rv{\textbf{Different online platforms offered unique interaction styles and community characteristics.} Commonly mentioned platforms were DevForum (Roblox official community), Discord, Reddit, and Korea-specific platforms like Naver Cafe and KaKaoTalk, as depicted in Fig. ~\ref{fig:communities}. These communities revealed dramatically different characteristics that profoundly shaped user experiences based on community ownership, modality, and culture. For example, DevForum is a structured}, text-based platform for learning, question-asking, and solution searching. Moderated by Roblox, it had more limitations on what could be shared \rv{due to} safety precautions. In contrast, social-media-based developer communities such as Discord presented a more dynamic environment with more functions.
\begin{quote}
    \textit{[Roblox-related Discord servers] are friendly, more energetic, and a little bit more chaotic but have more freedom, and flexibility and don't have as stiff an atmosphere as DevForum. (C17)}
\end{quote}

\rv{This was possible because Discord supported more interactivity between users through features such as voice chats and screen shares. Most importantly, anyone could create their own community servers and decide the rules. Participants mention sharing funny memes as a joke, playing games together in independent voice chats with community friends, or showing their live coding progress to community members like a ``code with me (C9)'' in peer-led communities. To many participants, Discord communities were preferred for the ``real-time communication (C10)'', and the ability to start your own community easily}. \\
 
\rv{\textbf{Participants strategically navigated multiple communities to meet diverse personal and professional needs.}} As teen developers explored and adapted to different communities, they began strategically using these digital landscapes. \rv{As C13 explained, their approach shifted from widespread engagement to more purposeful targeted participation:}
\begin{quote}
    \rv{\textit{Twitter is where Roblox employees can easily discover developers by looking at their work, while Naver cafes are preferred by professional companies. Twitter allows light self-promotion and easy visibility. Discord enables real-time work sharing. As I developed deeper relationships in Discord developer communities, I realized uploading my creations to every community was inefficient, so now I focus on Discord communities only. (C13)}}
\end{quote}
 
\rv{Korean teen developers like C13 used common social platforms in Korea including Naver Cafes and KaKaoTalk Groups, which also served as a first step into other global developer communities. C14 first learned about Discord after talking with members of KaKaoTalk Groups and Naver Cafes and got to know other servers after being familiar with Discord. However, for most participants, including C13 and C14, DevForum and Discord were reported to be by far the most frequently used, so our analysis focuses primarily on these two platforms.} 
% \rv{(transition sentence. For Korean teenager developers, the Naver Cafe and KakaoTalk served to introduce them as the first step into the developer communities. One QUOTE ). Our analysis focused primarily on DevForum and Discord due to their overwhelmingly dominant usage among participants. 

\begin{figure}
    \centering
    \includegraphics[width=1.0\linewidth]{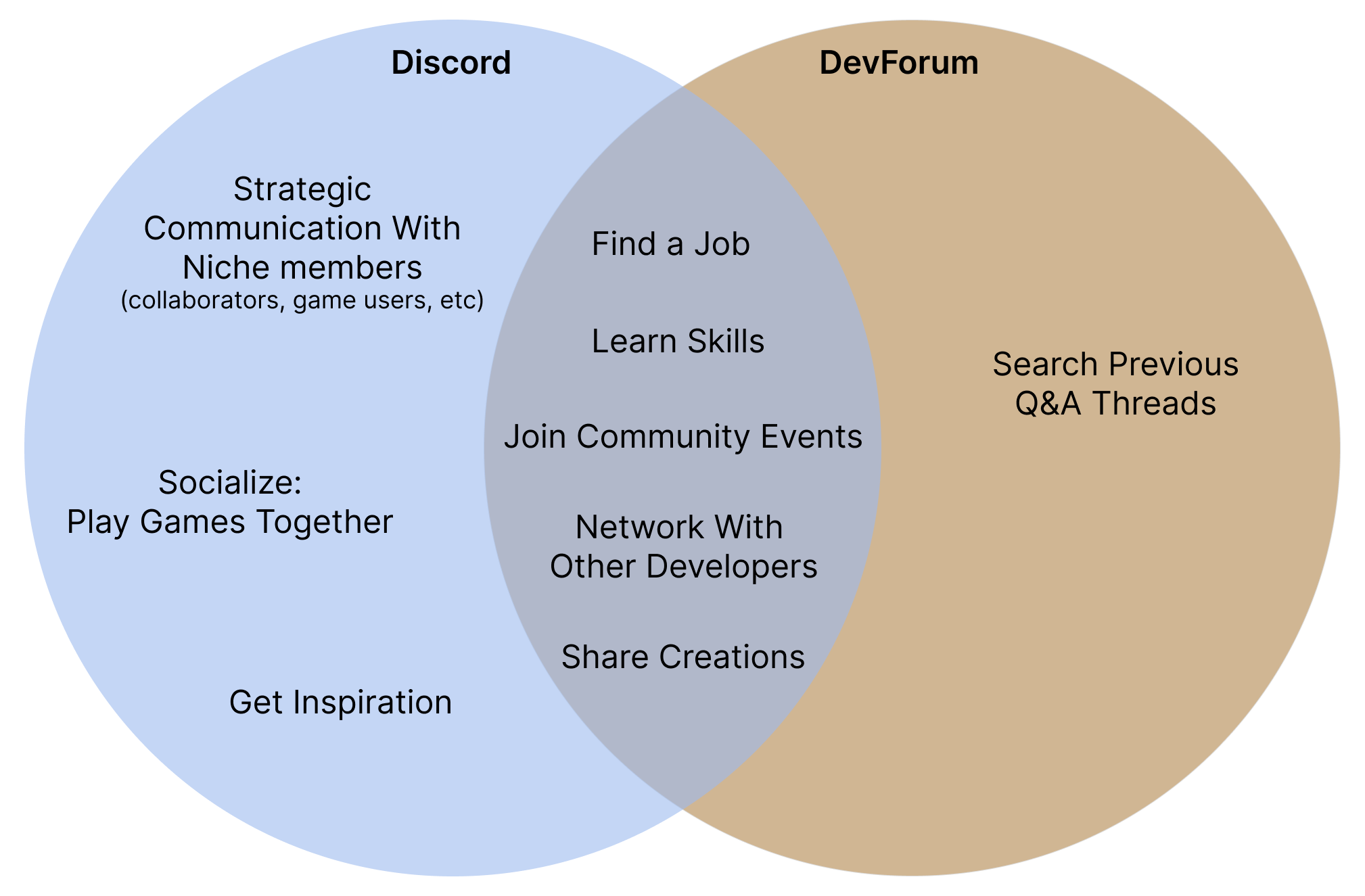}
    \caption{Comparison of online developer communities. We only visualize the most commonly used with at least one differing usage.}
    \label{fig:usage}
    \Description{\rv{(Figure updated)}A ben-diagram that compares Discord and DevForum, each represented by circles. The figure shows the common usage purposes in the overlap of two circles, and the remaining parts list the unique usage purposes of each community.}
\end{figure}

\rv{These platforms illustrated how seemingly similar activities could manifest entirely differently across digital spaces. For instance, the simple act of sharing game creations took on distinct characteristics depending on the platform. On DevForum, sharing meant formal presentations with detailed technical explanations and structured feedback. In contrast, Discord communities transformed the same activity into a casual, real-time interaction—developers might quickly screen share their work, get immediate reactions, and engage in spontaneous collaborative discussions.} The varying usage is depicted in Fig. ~\ref{fig:usage}.

\rv{Teen developers visited DevForum for formal questions and to access comprehensive game development resources (e.g., the Q\&A database). In contrast, participants used Discord as a decentralized ecosystem where teen developers could create their own communities. C11, for instance, explained this diversity of communities, noting~\textit{``I used one for scripting help, another for finding jobs.''} C18's approach exemplified this strategic community usage, participating in over 200 Discord servers to gather a diversity of inspirations:~\textit{``I get ideas from community discussions across platforms. I noticed kids learning grammar, so I created a game that improves English skills by using GPT to generate random word explanations.''} Some participants created their own Discord communities to connect with their games' players (C7), their YouTube followers (C16), or their UGC buyers to get feedback on their new 3D models (C4). C4 mentioned how he realized his UGC was not as popular as he thought it would be in the market, compared to the efforts he made. He initially decided to hear from other developers who were his customers on what features to add, which led him to make his own Discord community with developers which eventually reached a size of 2700 members at the time of the interview.} 

\rv{This diversity of approaches shows how teen developers don't just use communities—they strategically construct digital ecosystems that support their learning, creativity, and professional growth by understanding and leveraging the unique social dynamics of each platform.}

\subsection{Benefits from participating in developer communities (RQ2)}
This section outlines the benefits reported by interviewees from their participation in online developer communities. \rv{Participants enthusiastically reported the benefits that each of their communities could support.}

\subsubsection{\rv{\textbf{Learning and improving technical skills.}}} 
%J: I think we want to go into significantly more depth here to identify the breadth of skills and how the community helped develop them.

%\textbf{Self-directed and Collaborative Learning}
\rv{When participants first joined developer communities, they often joined for the purpose of asking very specific questions such as \textit{"Why does my structure keep collapsing? How do I fix this phrase error?"} Our interviews resoundingly confirmed that learning technical skills to create a game or design game elements was the most prevalent benefit, being the initial motivator for participating in the communities in the first place. } 
%Learning technical skills to create a game or element was the most prevalent benefit echoed by almost all participants. 
Such learning typically emerged from searching for related questions \rv{that had previously been} answered or asking \rv{new questions} to the Q\&A channel within the community. \rv{C4 shared that in the beginning, he asked a lot of questions to learn how to use the Roblox Studio UI and learn answers that didn't show up when searched.}%New creators frequently asked questions about bugs or problems they encountered and received solutions from the community. 

Experienced \rv{Roblox} developers, with over five years in their profession, continued to seek advice from community members to resolve complex issues in development. C10 mentioned one of his communities having ``truly skilled" developers well beyond the level of hobbyists, whose advice was like a guaranteed map to resolve whatever problems \rv{community members} were having. Even participants who were usually passive observers gained valuable insights, such as improved 3D modeling techniques or new mathematical concepts for animation \rv{by looking at the active community-level Q\&A channels}.  %\textcolor{white}{space}\\
\begin{quote}
    \rv{\textit{I have learned a lot of stuff. I [wouldn]'t be good like this to make my own games. If DevForum didn't exist, then I couldn't solve problems that I struggled [with]... that forum is really important.}} (C18)
\end{quote}

    %\textit{In the beginning, I asked a lot of questions. The Roblox Studio UI can be daunting at first time. For example, I was wondering ``Why does my structure keep collapsing?" I ended up asking people a lot of questions like this that don't show in searching.} (C4)}
    
\rv{Many of our participants also described learning from collaborative team projects they joined within the community. Participants would find opportunities to join a group for game development in the channels dedicated to hiring collaborators (section 4.1). Working in small groups was} another big part of \rv{the learning process} as participants often discovered their preferred roles (e.g., game design, scripting) while trying out different team dynamics. Collaborating with more experienced developers facilitated skill development, as experienced team developers would give advice and tips on game development including file organization or shortcuts. \rv{Participants like C12 mentioned how team collaborations would also build relationships making it easier to ask questions in a 1:1 manner instead of publicly asking within the whole community.}
%\textcolor{white}{space}\\

\begin{quote}
\rv{\textit{My collaborators, friends actually, are really amazing. They even helped me build a map once. I ask them a lot of questions when I'm stuck, \rv{too much to post in the official Q\&A}. And sometimes I did help them too once - very proud of that. (C12)}}
\end{quote}

\rv{Additionally, shared} community resources \rv{supported participants' self-directed learning.  C6 acknowledged how using free 3D assets helped him transition from beginner-level creations to complex, customized designs.} Sophisticated 3D models, music, codes, \rv{Roblox Studio updates,} and AI tips \rv{were shared} within the communities - some having separate channels for sharing. These helped participants \rv{to stay updated, quickly build games, and later} explore how other community members structured their code or composed 3D models: 

\begin{quote}
\rv{\textit{I feel around 40\% of the stuff I learned was from the communities. The resources really help you \rv{because it's daunting to build everything at first.}. People just share [the resources] kindly. (C6)}}
\end{quote}

%Updates on new Roblox features or bugs shared \rv{in the community} also helped \rv{participants} stay informed and build more efficient projects. 
\rv{Participants like C9 mentioned how community} feedback channels \rv{further} motivated participants to continue their effort \rv{to overcome challenges. Consistent with Roque et al. arguing that collaborative learning happens in online communities~\cite{roque2016children}, we find that developer communities fostered learning as a community in practice. Further, learning for participants was} a fun and self-motivated activity \rv{supported by their participation in developer communities.}

\subsubsection{\textbf{Developing social skills by building networks and relationships.}}
\rv{Developer communities weren't only focused on building games. Often they also dedicated space for building networks and relationships with other members. As the communities fostered interaction between users to discuss and collaborate,} participants could develop soft skills through this online communication. \rv{For many participants, the space felt like an extension of their social lives or even their full social life -- in the case of a participant who was home-schooled -- particularly because Roblox communities were unique.} Unlike other developer communities, Roblox’s collaborative and ``peer-led'' environment, \rv{especially on Discord,} is characterized by members of similar ages and informal interactions. Participants mentioned they could \rv{guess} community members were young by the jokes they shared and \rv{by how} the time people showed up seemed \rv{to be} after school. This sense of camaraderie and shared interests fostered a unique ecosystem where participants felt at ease discussing development, collaborating on projects, and communicating in a more \rv{casual and fun-driven way. C7 described his communities as a space where people like him could breathe:}.
\begin{quote}
    \rv{\textit{I really enjoy that kids my age can talk about \rv{(game)} development stuff. In places like Unity or elsewhere, most of the people are older, busy working, there's not really room for saying ``Hey, I want to make a fun game like this, let's do it together''. The conversations are more serious, about profits rather than fun. The atmosphere is entirely different. (C7)}}
\end{quote}

 %This space made the community atmosphere relaxed and fun for users which contrasted with the more serious and profit-focused conversations found on other platforms like Unity. 
 
 %\textcolor{white}{space}\\

\rv{But working with others online was not always easy.  Self-motivated collaborations with group members sometimes encountered challenges when team members had different goals for development (e.g., for fun, for the money), and as such they required tact, patience, and openness.} Participants learned to navigate interactions with diverse cultures like C9 who mentioned having worked with friends on three continents. This expansion beyond their local network meant they had to adapt to communicating online. \rv{Participants mentioned the communication strategies they have learned through the process of collaboration -- starting conversations kindly, being clear about intentions, and finding common ground. }
\begin{quote}
    \rv{\textit{I now add labels to my 3D models so the scriptors understand what I mean. That really helped us sync (C2).}}
\end{quote}

%This type of environment facilitated self-motivated collaborations, making effective communication essential. 
%For instance, C2 learned to communicate their intentions clearly by adding labels to 3D models to better align with scriptors. 

\rv{As their confidence and self-efficacy grew, }participants \rv{began taking proactive roles within communities. Participants like C12 mentioned they gained self-assurance} in promoting their skills in the community. Some participants \rv{also} mentioned taking leadership roles within projects as they gained more confidence in how to manage talking within the community. Four \rv{participants (C2, C4, C6, C9} took moderation roles in Discord developer communities where they once joined as newcomers. Some participants forged friendships and \rv{developed social capital, improving their reputation within the broader community.} Some participants (C3, C7) who did a few projects together discovered they lived close \rv{to each other} and managed to meet and become offline friends. For instance, participants reported that these casual local meetups provided more personalized interactions, which proved to be important for building long-term relationships. This confidence from community involvement led them to explore new roles, such as creating Discord communities or expanding their activities on platforms like YouTube.\footnote{\url{https://www.youtube.com/watch?v=lnhrFUmFpak}. C6 mentioned creating his YouTube channel to share his mini-games with a broader audience.} Three participants (C1, C4, C17) built their own Discord developer communities to engage with players of their game, collaborate with employees of Roblox, \rv{with C6 seeking to} establish the largest Roblox developer community in Korea. These findings show \rv{ how Wenger's concept of Communities of Practice} succeeded naturally in teen developer communities, where participants gradually took different roles and identities in the community~\cite{li2009evolution}. Most participants mentioned recognizing the importance of community mentoring and were willing to pay back the help they got from the community.

%also improved their promotional skills and gained confidence in organizing and presenting their work to be selected for a hiring post in the community. Through these collaborations, they forged friendships and expanded their social networks.
\begin{quote}
    \rv{\textit{At first, I was just an observer. Now, I'm leading a game I thought of and having a voice in the community. It's amazing to be recognized for my skills and be helpful to others. (C14)}}
\end{quote}

\rv{Within a community built on shared ambitions and playful collaborations, participants could feel heard and, through this connection, experience a sense of belonging.} %Broadly, participants developed effective online communication skills and built meaningful relationships within the community.

\subsubsection{\textbf{Turning Hobbies into Careers.}}
%External
%Starting Point to expand
%Hobby --> to Work, Job 
%As participants became more advanced, they would often find themselves being the replier than the questioner.

\rv{For some participants, Roblox development was a hobby. For others, it became something far bigger. Monetization opportunities, such as commissions from collaborations and revenue from creations, served as a strong motivator to continue development for those who consider Roblox development beyond a hobby.} 13 participants pursued commissions from community hiring posts, creating a vibrant marketplace for part-time work. As such, C13 sought in-community jobs for income, \rv{which was} particularly beneficial in countries like Malaysia with favorable exchange rates for dollars earned, while C17 earned money from selling his UGCs (avatar items), keeping him motivated. Community events with prize rewards further enhanced these financial incentives. \rv{Four participants had made revenue from their developed games, and two participants had worked together on the same very successful game, which had temporarily been among the top 10 most played games on Roblox. While one teen developer made a game from his own idea, another participant (C4) mentioned he was invited to the team, and suddenly the game ``blew up''. This kind of success of turning developers' games into huge money was often a common story that participants were aware of and strove for themselves, working in a form of aspirational labor as described above.} While "fun" was the main motivation for creating Roblox games, monetization provided an \rv{overall} substantial additional boost for \rv{nearly all participants.}

\begin{quote}
    \rv{\textit{I get paid a lot by Roblox simply because people like to play my team's game.  After members saw my work, I was invited to join the team community. And when you get paid, that kind of just makes you more intrigued and keeps me developing. Also, I learned economics that way. (C3)}}
\end{quote}

%Monetization in Roblox significantly motivated participants by offering financial incentives. Four participants earned revenue from collaborative projects and engaged in revenue-sharing discussions on Discord. 

%\textcolor{white}{space}\\

\rv{As participants grew more skilled}, creating different projects (see Fig. ~\ref{fig:creation}), recognition within their communities followed. \rv{Some participants, such as C16, were recognized as official Roblox ambassadors, and such participants had chances to go to local Roblox conferences, meet senior developers, and develop strong social capital. C16 mentioned having TikTok and YouTube followers help gain him recognition, though he thought his game development skills still need improvement. From these experiences, C16 dreamed bigger: going to the world Roblox conference and pitching his game.} \rv{Skill enhancement led four} participants \rv{(C1, C2, C6, C10))} to secure freelance work \rv{under a formal contract with a small third-party company} and to secure longer-term employment in a few cases\rv{(C13))}. Despite lower initial offers given due to \rv{their} age, their proven skills \rv{eventually} led to better opportunities and, for some, serious consideration of a career in Roblox development such as a game developer or 3d modeling designer. Some saw their hobby as a potential career path and pursued related opportunities in \rv{the form of} a full-time job or related majors in school. \rv{For example,} C5 planned to go to a game-development-oriented high school. \rv{Similarly, C7 mentioned her career growth:}
 
 %such as C14, who became a mentor and attracted attention from Roblox staff on the DevForum. Others took on ambassador roles, promoting the community online and at events like the annual Roblox developer conference. 
 
\begin{quote}
    \rv{\textit{Actually, at the company I’m with now, I’m making my own game. I joined this company because they made me a good offer in DM, saying I could continue my side project and still receive a regular salary. The only thing is, I have to give a portion of the earnings to them. (C7)}}
\end{quote}

\rv{While some participants didn't view Roblox development as a long-term path—three mentioned shifting focus toward school or other pursuits—15 participants valued the foundational skills they acquired, such as understanding 'if' statements and loops, and felt confident applying these abilities in other fields. Contrary to prior research suggesting that Roblox creates a lock-in ecology for developers~\cite{kou2024ecology}, seven participants had positive experiences, discovering a passion for problem-solving or design, and expressed a desire to continue developing their skills beyond Roblox. This includes learning advanced modeling software like Blender and exploring other programming languages such as JavaScript (C15).}

%, such as creating Discord communities or expanding on platforms like YouTube\footnote{\url{https://www.youtube.com/watch?v=lnhrFUmFpak}}. Three participants built their own Discord developer communities to engage with players, collaborate with employees, or establish the largest Roblox developer community in Korea. 

\begin{quote}
    \rv{\textit{I'm pretty confident with Blender\footnote{3D modeling software not made by Roblox but often used among Roblox developers for more complicated modeling} now. I've noticed people using UnReal in Unity games at my current company. I'll probably try it or 3DMax (C2).}}
\end{quote}
    %\textit{Making games and the Roblox community was helpful; basics like if statements and loops can transfer to other programming languages (C17).}\\
    %\textit{I talked about which high school I should go to with a senior friend, and he recommended some game development-oriented schools which I agreed (C5).}\\
    %\textit{I discovered I like problem-solving. I would like to also learn other languages like JavaScript next (C15).}\\

\subsection{Challenges from participating developer communities and \rv{associated} coping strategies (RQ3)}

In this section, we report the \rv{various} challenges participants \rv{experienced in developer communities related to community accessibility, sustainability, and inter-user challenges between members.} 

\subsubsection{\textbf{\rv{Struggling to find the right community}}}
\rv{Although} online communities offered significant benefits for growth, \rv{many participants faced difficulties discovering and joining the right ones.} \rv{Discord, widely used for its flexibility in creating community channels and having diverse modes of communication, was the preferred platform for most developer communities. However, its invitation-only model often created barriers for newcomers. For instance, beginners like C8 were unaware of these communities until being introduced to them during our interview, revealing the challenges participants face in accessing some online spaces.} According to moderator participants, \rv{restricted access was necessary to ensure safety,} protecting \rv{community members} from \rv{potentially-problematic users}. 
%Some beginners like C8 were unaware of these communities and expressed interest only after learning about them in interviews, highlighting the challenge of balancing recruitment with safety, especially in smaller, heavily moderated groups. 
%But many were hard to \rv{find and join with accessible only through invitation policies. 

%Discord-based communities were popular among participants,  find due to invitation-only policies. 

\rv{Once participants joined a community, integration into the community posed another challenge.} While this process was vital for their continued involvement in the community~\cite{freeman2019exploring}, navigating Roblox-specific terminology and community dynamics \rv{felt} overwhelming,  
\rv{especially for young participants joining their first online communities}. \rv{The diversity of communities meant that experiences varied widely, with some being welcoming and others less so. C16, for example, described feeling isolated when his questions were ignored or met with unhelpful judgmental responses without explanation, which left him discouraged and questioning his place in the community. }
\begin{quote}
    \rv{
    \textit{I reached out to one of my collaborators once. They solved it [my problem] but didn't explain it to me. Instead, they said my code was really messy which it isn't (C12).}}
\end{quote}

\rv{To get used to the community, participants mentioned various strategies such as trying to spend more time, observing others, making friends through collaboration projects, or playing games together with community members.} Advanced teen developers faced challenges in finding the right \rv{communities that matched their skills and offered meaningful} growth and networking opportunities. \rv{Sometimes, the communities they belonged to felt too small for them or did not have enough resources for them to grow. Language barriers also created obstacles. While} many \rv{prominent} developer communities were English-speaking, non-native speakers \rv{from Korea, Japan, and Spain, often felt more comfortable in communities that used their native language.} However, \rv{such} communities \rv{were harder to find. \rvtwo{For instance, when C4 wanted to access more advanced developer resources, he found no materials available in Korean and had to resort to translating English resources.}}

%\begin{quote}
    %\rv{\textit{To be honest, there aren't a lot of sources in Korean. Roblox Korea started recently, which is better, but still, the advanced resource information is hard to find [in Korean]. So, I use translators. (C4)}}
%\end{quote}
%\begin{quote}
    %\rv{\textit{I think Japanese [Roblox] developers could have better opportunities if they knew English (C16).}}
%\end{quote}
\begin{quote}
    \rv{\textit{I have worked with a lot of different people - American, UK, Brazil, and other EUs. Personally, I like Spanish people the best as a Spanish creator. But, from what I know, there aren't Spanish Roblox Creator communities out there. (C14).}}
\end{quote}

\subsubsection{\textbf{\rv{Struggling to Balance Commitments.}}} \rv{Balancing Roblox development} with academic commitments \rv{emerged as a major challenge for teen developers}. Participants who \rv{had} used the community for a long time, such as C14, \rv{reported} that many peers became inactive as they \rv{advanced in school, disrupting} community dynamics and \rv{slowing} down the progress of collaborative projects. \rv{This was especially challenging when a knowledgeable mentor had to leave the community. C15, who enjoyed helping out and answering questions in his communities, felt it becoming harder and harder as he was doing it voluntarily and faced more and more serious school works as an 11th-grade student, but he was concerned about the impact his absence might have.}
%\textcolor{white}{space}\\

\begin{quote}
    \rv{
    \textit{Well, there are three large times when people disappear -- starting middle school, high school, and college admission. Some do come back, but others don't, especially if they decide to focus on studying. (C7)}
    }
\end{quote}

Participants who continued in Roblox development often \rv{aimed to pursue it as their career, sometimes at the expense of their education.} For instance, one participant (C13) \rv{became so} overwhelmed and very sleep-deprived \rv{with balancing her Roblox projects and school work that she took a semester off from school to concentrate on her Roblox job} \rv{creating new 3D} maps and UGCs. Others viewed Roblox as a hobby, balancing it with their education, like C18, who chose to \rv{leave} Roblox \rv{entirely} to focus on his college plans.

\rv{Social} perception of Roblox as ``childish'' also \rv{further complicated the sustainability of the communities.} As teenagers, participants were \rv{conscious} of how \rv{their involvement was perceived negatively by} parents, teachers, and peers outside of Roblox, which sometimes undermined their motivation. Participants like C7 mentioned facing opposition from his parents, school friends, and teachers because Roblox \rv{development was perceived as \textit{immature}} to pursue as a high school student.

%\textit{There was actually a lot of opposition from my parents. There was a strong perception that this (Roblox) was for kids, and none of the students at my school were doing it. My friends around school or the teachers didn't think positively about it either. I had to prove myself. (C7)}\\

Many participants, aware of these perceptions, initially concealed their \rv{activities} in Roblox development. For example, C13 described hiding her activities until she achieved significant milestones, \rv{which felt important enough to gain validation from external people.} 

%others viewed them, the negative image sometimes undermined their motivation.
%impacted how participants were viewed by outsiders, including parents and teachers. 

%\textcolor{white}{space}\\

\begin{quote}
    \rv{
    \textit{Oh wow. It felt like living a double life - a secret small hobby ... until I reached a certain level of quality in creating what I imagined. Now, my parents show my work to relatives. They usually say it’s fascinating or impressive. (C13)}
    }
\end{quote}

\rv{Some participants eventually won their parents' support by demonstrating the financial potential of their development. Monetization often emerged as a validation--proof that their work mattered. Like prior work on TikTok teen creators~\cite{bulley2024dual}, making money convinced their parents to believe in their activities. As in the case of C6, winning a Roblox competition or earning game income helped skeptics see Roblox development as a legitimate career path.} This monetization, however, brought its own set of challenges, which will be detailed in more depth in the Discussion~\ref{discussion:monetization}. 

\begin{quote}
    \rv{
    \textit{At first, my parents were against it, but I did it anyway. They said like, `` you're spending too much time on something unproductive.'' But once I won some Roblox challenges prize and received strong revenue, it helped them see Roblox's potential for a serious career. (C6)}
    }
\end{quote}

%defining moment for participants thinking of a serious career in game development. 

%A few participants mentioned successfully gaining parental support for Roblox development by proving their revenue and creations.  
%\textcolor{white}{space}\\

%    \textit{There was actually a lot of opposition from my parents. There was a strong perception that this (Roblox) was for kids, and none of the students at my school were doing it. My friends around school or the teachers didn't think positively about it either. I had to prove myself. (C7)}\\

\rv{In rare cases, participants} attended schools that \rv{supported game development, integrating} Roblox development into their education. These students benefited from taking courses on programming or 3D design, \rv{which they applied} to their creations. \rvtwo{Some schools actively promoted Roblox development, with C16 describing how it was encouraged by his science teacher and even featured in school-wide events organized by the student council.} Participants like C4 had the resources to consult with \rv{school} teachers for advice, \rvtwo{including valuable input from specialized faculty such as architecture teachers who provided guidance on creating immersive spaces. This created} a synergy between their schoolwork and development efforts. \rvtwo{For instance, C7 chose Roblox as a career path after exploring other platforms in school. Beyond formal education, some participants found support through private instruction or family connections - C12's parents invested in private Roblox coding lessons, while C17 received guidance from family members who worked as programmers.}

%This synergy helped C7 to choose to focus on Roblox as a career path. One participant (C12) had supportive parents who paid for private Roblox coding lessons, and C17 had programmer family members who helped him out occasionally. 

%\rv{For C16, Roblox development was encouraged in his programming class, so he had offline friends to talk with about Roblox development. }
%Among those who gained parental support, some 
\begin{quote}
    \rv{\textit{I go to a school that specializes in game development. They mainly teach tools like Unity or Unreal Engine, and we release game on Steam or Google Play. Roblox, however, felt much friendlier and better for monetization. So, I looked into it more seriously. (C7)}}
\end{quote}
%\begin{quote}
    %\rv{\textit{I sometimes talk to my homeroom teacher about my (Roblox) games. Since he's an architecture teacher, his advice on creating immersive space really helps. (C4)}}
%\end{quote}
%\begin{quote}
    %\rv{\textit{Roblox was a big thing in my school. We had a creators' challenge arranged by the student council, and my science teacher encouraged us. (C16)}}
%\end{quote}

\rv{Despite these positive examples with offline resources, participants were discouraged by negative perceptions saying that there were not many resources available outside online developer communities for a longer term career, which led} many participants \rv{away from} game development, impacting the sustainability of their communities as members came and left. 

\rv{In response to the challenges of community engagement,} many beginner developers turned to AI tools, mostly ChatGPT, for \rv{support}. \rv{Participants} found AI to be convenient \rv{and effective for basic coding tasks, idea generation, and design feedback without feeling like they were burdening community members. However, most }agreed that AI alone could not \rv{replace the depth of knowledge and collaboration found in communities. Complex issues still required human insight and collective problem-solving.}
%address complex levels. Therefore, while AI was useful for basic queries, it continued to rely on community engagement for more intricate issues and collaborations.

\subsubsection{\textbf{Dealing with financial scams}}

\rv{Participants claimed} both financial scams and inter-user conflicts as significant challenges within developer communities. \rv{Consistent with findings on harmful game design in previous research}~\cite{kou2023harmful}, financial scams \rv{were a recurring issue. Participants either personally experienced} being scammed \rv{or knew} others \rv{who had fallen victim}. Scams often \rv{included developers} being underpaid or not paid at all, even \rv{for} successful \rv{projects}. In some cases, \rv{participants paid in advance for work that was never} completed. C18 \rv{reflected on how} the frequent collaborations that happen in the communities made reputation and trust crucial \rv{for successful projects. This trust-dependent nature of collaborations often left} new members \rv{vulnerable to exploitation:} 
\begin{quote}
    \rv{
    \textit{You should like be aware of people because sometimes some young developers trust other developers too much and get scammed. (C18)}
    }
\end{quote}

While scams were less common in \rv{developer}-focused communities \rv{than in} Roblox player communities trading UGCs, participants \rv{still viewed} them as prevalent in certain developer communities \rv{among members especially those who want to make profits}. \rv{Moderator participants explained that} scammers often exploited community norms of mutual trust and \rv{conducted fraudulent} activities in private chats, beyond the reach of moderators, and also bypassed bans by creating new accounts. Six participants \rv{explained how} this left the scams to be addressed by the young community members themselves. \rv{Even in cases of blatant plagiarism or impersonation,} responses varied \rv{widely.} For example, one participant (C18) mentioned \rv{that he felt that it was acceptable for others to copy his game}, as \rv{he felt that it would eventually lead to more growth of his game through the popularity of similar games}. Other participants \rv{took creative or legal steps to protect their work.} For example, C5 mentioned trying to resolve this by learning about \rv{intellectual property law and norms}, which was not covered by \rv{her} school education, \rv{and as a result she developed better} watermarking that could not be deleted easily. \rv{Another participant (C9) said he} sabotag\rv{ed} a collaborator by inserting infinite loops into the code to address non-payment. 

\rv{However, dealing with plagiarism and financial disputes often felt overwhelming, especially for younger participants. Most collaborations proceeded in an informal way between community members without any written contracts. Except for the few participants employed by companies and working under a formal, legal contract, the legal aspects were described as a confusing and unfamiliar concept, leaving both the participants and their legal guardian(s) unaware of how to navigate such matters. C2, for example,} described feeling powerless in a case where their development team earned only a fraction of the revenue generated by their game:
\begin{quote}
    \rv{
    \textit{Our developer team received about 2,000 Robux a month\footnote{Approximately US \$25 at the time of writing.}—when we know the game is making 200,000 to 500,000 Robux monthly. We didn't report it because it was hard to find the legal documents. Also, within Roblox, there is an implicit atmosphere of just trusting each other. The private matters usually stay between the individuals and not the whole community. So, as I developed a lot of games, I experienced not being paid, or someone using hacks on my game, copying it, and putting it in their own game. (C2)}
    }
\end{quote}
\rv{Such experiences led many participants to be more cautious about collaborations, while some became demotivated toward creating games together. Despite the community’s collaborative spirit, the lack of formal protections left many young developers vulnerable.}

\subsubsection{\rv{Dealing with Inappropriate Users.}}
\rv{Although Roblox creator communities were widely perceived by participants as safer and more respectful than Roblox player communities, instances of inappropriate behavior were not entirely absent.} Trolling, disputes over critical feedback, and unauthorized self-promotion \rv{emerged as recurring challenges that participants consistently confronted. More alarming incidents included egregious violations such as banned users disseminating explicit AI-generated content. Female participants mentioned seeing uncomfortable chats that made them choose not to use their voice-chats and to use masculine-looking avatars to conceal their gender. In line with prior work \cite{kou2024ecology} some participants experienced situations where other developers promoted games embedded with radical nationalist ideologies. While they were banned from their community, the problematic users made new accounts again and continued to create these games.}
%In the Roblox creator communities, inappropriate behavior occasionally surfaced. 

%Participants felt that these environments were generally considered cleaner than Roblox player communities, as members' focus on collaboration and reputation fostered a more respectful atmosphere, but some exceptions persisted. More common instances of inappropriate behavior included trolling, disputes over negative feedback, minor rule evasion, and unauthorized self-promotion, but participants also reported more serious safety concerns, such as a case where a banned user shared explicit AI-generated links and another case where a female creator felt uncomfortable with sexual comments made by male creators. Participants also mentioned rare cases of community members promoting games with historically problematic ideologies on radical nationalism.

\rv{The likelihood of encountering such incidents often depended more on how a community was managed than its size. Participants claimed to prefer larger developer servers, as they had dedicated moderation teams and clear rules, typically providing safer spaces. In contrast, smaller, close-knit communities, built on presumed mutual understanding and personal connections, sometimes inadvertently created environments where problematic behaviors were systematically overlooked. This inconsistency shows how effective moderation practices played a more decisive role in shaping the community atmosphere than member count. This led to C10, a community manager,} having an invitation-only approach to maintain a respectful environment. 

\begin{quote}
    \rv{
    \textit{I manage communities where all the members are verified [with a size of 300 users]. In our server, we try to maintain a clean environment with good developers. There’s a lot of constructive conversation between the admins and the users. But sometimes people from the shady side also join, so we have to deal with them occasionally. Because of these issues, we don't promote our community. (C10)}
    }
\end{quote}

\rv{The occasional presence of inappropriate users underscored the ongoing challenge of cultivating trust and professionalism in Roblox developer communities. These incidents reminded the participants of the delicate balance required to sustain a positive and inclusive environment for collaboration.}

\section{Discussion}
%1. Polishing, shortening
%2. Motivating - weak version vs persuasive --> We need to develop better for young people
%Sign posting throughout the paper. We need this information. Reminding in small pieces. 

\rv{Our findings have} identified \rv{that participation in Roblox developer communities fostered individual growth among teen developers, particularly in building technical skills and social communication skills. These benefits align with prior work on indie game developers, where collaborative environments encourage democratic leadership and shared ownership of projects~\cite{10.1145/3375184}.} However, these social and \rv{technical} benefits \rv{coexist with} challenges \rv{such as financial scams and interpersonal conflicts. This duality emphasizes the need to take a more comprehensive view:} what makes a positive, safe online \rv{community} for teen \rv{game developers}? \rv{In this section, we address this question by highlighting key components of growth-fostering online communities, complexities introduced by monetization, and implications for fostering safe environments for teen developers.}

\subsection{Key components for fostering growth in online developer communities}
%\begin{tcolorbox}[colback=gray!10, colframe=gray!50, width=\textwidth, sharp corners=south, boxrule=0.5mm, title=]
%\textbf{What are the key features in providing safe teenager-friendly online space that fosters growth?}
%\end{tcolorbox}

In this section, we identify three \rv{critical} factors driving many of the benefits \rv{participants reported from} online \rv{Roblox} developer communities\rv{: play as a core motivation, opportunities for meaningful communication, and access to growth-enabling resources. Some of these} factors were strongly aligned with previous work but manifested in new forms in the in-the-wild, play-driven case of Roblox developer communities. 
\subsubsection{{\textbf{Self-driven play as a Core motivation.}}}
Teen developers were self-motivated by \rv{the intrinsic enjoyment of creating and experimenting within the Roblox ecosystem, reporting fun as their primary factor for why they develop games. This demonstrates how playfulness remains a cornerstone of sustained community engagement} and growth in teen-created and moderated spaces. \rv{Similar findings} in developer communities like Scratch \rv{show} technical development \rv{gained through collaborative learning in communities} ~\cite{brennan2021kids, 10.1145/3491102.3502124, roque2016children} or adult-focused platforms like StackOverflow~\cite{10.1145/2531602.2531659, 8417152}. \rv{However, unlike Scratch, where play occurs within structured learning environments, Roblox supports a hybrid model where teens independently create and moderate their spaces. }This dynamic fosters a unique sense of autonomy and ownership, consistent with theories of participatory culture~\cite{jenkins2009confronting, jenkins2015participatory}.

\rv{Freeman highlights the critical role of small teams and "democratic" participation, wherein collaboration fosters shared decision-making and individualized contributions~\cite{freeman2019exploring}. Similarly, teen-led Roblox communities function as small, dynamic ecosystems where developers work collectively to design and iterate on games. Despite being self-organized, these communities mirror the socio-technological challenges faced by indie developers, especially in identifying the right collaborators with shared goals. For teens, these challenges manifest in managing their team formation and navigating technical systems with limited professional guidance.}
%fostered by peer learning and technical development in teen-created and moderated spaces, even without strong adult supervision. These benefits align with education-based developer communities ~\cite{brennan2021kids, 10.1145/3491102.3502124, roque2016children} or adult-focused platforms like StackOverflow~\cite{10.1145/2531602.2531659, 8417152}. %Our findings reveal that the benefits of participatory culture~\cite{jenkins2015participatory} of teen autonomy thrived even outside formal educational settings. 
\rv{Building on Freeman's insights, we argue that the playful roots of the participatory culture within Roblox developer communities are crucial for encouraging long-term participation and growth. Even as} challenges \rv{arise, such as encountering scams or interpersonal conflicts, the ability to experiment freely within} the collaborative and friendly atmosphere of Roblox developer communities enables resilience and creative exploration. 
%participants are drawn to the collaborative and friendly atmosphere of Roblox developer communities, which encourages creativity and trying out for fun. Building on previous research, we argue that maintaining a playful environment is essential for sustaining long-term engagement among teenage developers.

%\rv{Reviewer comment on including Guo's indie developer challenges, AI, personal growth with team. }
%\rv{Appropriate social, cultural, financial capital}
%\rv{Incentives play \& money }

\subsubsection{{\textbf{\rv{Access to rich technical and social resources facilitates learning.}}}}
\rv{Our findings underscore that access to both technical and social resources plays a crucial role in fostering growth within developer communities. For newcomers and even more experienced participants, the ability to connect with others is vital for learning, collaboration, and expanding their professional network. However, communication barriers often hinder this process. Without an environment conducive to meaningful interaction, some participants, especially beginners, turned to AI tools like GPT for technical support, particularly for coding tasks and idea generation. The increasing support-seeking from AI mirrors how indie developers use AI as their programming partner in idea generation~\cite{panchanadikar2024m}. \rv{However, while these} AI tools provided immediate assistance, they did not offer the depth of collaboration and mentorship that could be found in community-based interactions.}%However, unlike in Panchanadikar et al.'s work, our participants were not scared of career growth risks and were willing to use AI for their communities like organizing events.} %Opportunities for communication also emerged as an important factor. Newcomers and experienced participants all needed social connections within the community to learn, collaborate, and network. However, without a sufficient communication environment, some newcomers turned to AI tools like GPT for support and some more experienced participants avoided exploring new communities. Events, challenges, and local meetups played an important role in fostering interaction and overcoming barriers like language. Our findings suggest that beyond online communication, integrating offline, local-based gatherings could strengthen community bonds and provide a more personal connection for members.

\rv{Furthermore, we noticed resource differences beyond communities. Some participants felt that their communities had abundant resources to learn and had members who were active in answering questions, while other participants wanted to search for more advanced communities. This discrepancy suggests the importance of ensuring that all members, particularly beginners, have equitable access to the necessary tools, guidance, and networking opportunities that foster their personal and professional growth within the community.}

\subsubsection{{\textbf{Incentives can drive community growth.}}}

Lastly, providing incentives like monetization proved effective in fostering growth within communities, as seen when participants joined \rv{in response to} commission posts and developed their skills accordingly. \rv{Community-wide competitions with Robux prizes were also incentives for participation commonly enjoyed by participants.} However, \rv{these forms of} engagement \rv{were} voluntary, relying on the social value of the community to motivate participants like in other developer communities~\cite{lu2021motivation}. This made it challenging for advanced developers \rv{to participate, as they often struggled to balance community involvement with schoolwork and development loads.}
% struggled to balance community involvement with schoolwork and development loads. 
Thus, implementing \rv{formal, financial incentives and/or less formal, social incentives each can enhance growth in different ways for different populations.} \\

\subsection{The complexities of a monetized hobby ecosystem}
\label{discussion:monetization}
%\rv{To ADD monetization in other teen creator communities - TikTok, YouTube~\cite{lombana2020youth, bulley2024dual, golmgrein2023comprehensive}. The promise of potential future success motivates children to engage in unpaid creative work, a dynamic that can perpetuate systemic inequalities. Successful creators typically emerge from backgrounds with greater resources, access to technology, and supportive networks—creating a cycle that marginalizes creators from less privileged backgrounds.}\\

%Monetization isn't new - aspirational labor. But roblox differs in Xx.
\rv{Monetization motivating teen Roblox developers is not a new concept; it fits within the broader framework of \textbf{aspirational labor}, where teen content creators are motivated by the hope of future success~\cite{duffy2017not}. This phenomenon is well-established in scholarly literature, which discusses the developmental role of unpaid creative work and the intersection of passion, skill-building, and economic opportunity in youth participation built upon venture labor and hope labor models~\cite{neff2012venture, kuehn2013hope}. Roblox stands out from other content creation platforms, such as YouTube, blogging, and game streaming, because it combines game development with monetization—offering teens the opportunity to engage in both creativity and entrepreneurship; Unlike YouTube influencers or TikTok creators, whose success is often based on personal branding, game developers must create engaging, playable content, requiring specific technical and design skills~\cite{lombana2020youth}. Compared to many traditional game modding processes, Roblox Studio significantly lowers technical barriers, making it more accessible and empowering to teen developers. }%Monetization on platforms like Roblox represents a modern extension of aspirational labor, where young creators blend play with work, motivated by the hope of future success. This phenomenon aligns closely with scholarly discussions on the developmental role of unpaid creative labor and the interplay between passion, skill-building, and economic opportunity in youth participation. \textbf{Unlike teen content creators on platforms like TikTok or YouTube, Roblox developers operate within a structured creative ecosystem where technical skill and gameplay innovation take precedence over personal branding.} While YouTube or TikTok creators often monetize through advertising or sponsorships tied to their personalities, Roblox creators rely on crafting engaging, user-generated content (UGC), requiring technical expertise in game development. This distinction introduces unique challenges, as these teens must navigate complex systems of economic engagement and virtual economies with minimal oversight.}

%Like literature monetization clearly introduced more empowerment. With more easier barriers.
\rv{The appeal of Roblox’s monetization system is clear, as it enables teenage developers to make money from their passion. Many participants in our study reported that the platform offers them increased autonomy compared to other teen job options, such as part-time work, as it allows them to earn money independently with what is at its core a form of play, without direct parental oversight. Teen developers value the creative freedom Roblox provides, viewing it as a space to experiment, hone their skills, and build a reputation with the potential for success. Drawing on previous work by Postigo~\cite{postigo2010modding} and Taylor~\cite{taylor2009assemblage}, this mirrors the “modding” culture in the video game industry, but with more accessibility, diversity in making profits from game, game items, game assets, and technical skillsets. Some participants strategically leveraged their games to build social capital and professional networks, demonstrating the potential for both skill development and future monetary success.}

%However, it also introduced safety issues. As it blurs work and play ~~
\rv{However, the formal blurring of play and work through Roblox’s monetization system creates new challenges. While the prospect of financial independence is a strong motivator, the absence of formalized structures or clear guidelines on what constitutes “labor” leaves teen developers vulnerable. Participants described incidents of scams and exploitation within peer communities, highlighting their lack of protection and limited knowledge about navigating such risks. For teens in countries requiring parental consent for legal work, the situation was further complicated. Some participants who joined small companies faced significant hurdles due to their families’ unfamiliarity with the complexities of game development agreements. Many parents, unfamiliar with such arrangements—especially in the niche context of Roblox—struggled to judge the fairness or legality of contracts. This inexperience often led participants to accept vague terms, resulting in unclear expectations and under-compensated work. Two participants shared instances where what began as creative exploration turned into unpaid or underpaid obligations, consuming more time and energy than they had anticipated.} 

\rv{The monetization ecosystem itself can exacerbate these challenges, sometimes driving teens to prioritize profit over ethics. This pressure can lead to decisions that compromise the well-being of younger players, such as designing games with harmful elements to maximize profitability.} Furthermore, the financial burden associated with advertising and publishing UGC—such as the cost of publishing limited UGCs, which can reach 20,000 Robux (approximately 15 USD) per item—adds another layer of strain, as previously noted by Kou~\cite{kou2023harmful}.

\rv{Living in a digital landscape driven by entrepreneurial ideals, it is natural for teens to be drawn to the prospect of creating their own games and earning financial rewards. However, we noticed that among the participants, there was an almost unwavering belief that hard work could secure success—a sentiment that reflects meritocratic ideals. This firm belief can quietly erode well-being when the realities of inequality go unacknowledged~\cite{watkins2018digital}. Participants like C13 described his failure in market success solely primarily as personal shortcomings, pledging to work harder to overcome them. This type of belief system can turn a creative playground into a space of self-doubt and relentless striving~\cite{ross2012search}.}

\rv{Our findings also highlighted the disparities in access to resources that shape these success narratives. For example, one participant, C12, thrived with the support of parents who invested in his coding education and helped him navigate complex programming concepts, while others struggled due to a lack of support from their communities.  While our interviews cannot definitively determine how much these external resources contributed to success in game development, they underscore the importance of recognizing the roles that factors like mentorship, financial backing, and even sheer luck play in shaping outcomes.}

\rv{Therefore, to foster a healthier perspective on monetization and success, it is critical to help teens see the broader context of their aspirations. Sharing details of success narratives including finding the right team, the right endeavor, and the right collaborative support will lead teen developers to focus more broadly than on just profits. In this way, a more balanced approach to ambition can be cultivated without compromising the well-being of developers--one that empowers teens to pursue their goals with resilience while acknowledging the complexities of the digital marketplace.}

\subsection{Takeaways for fostering safe and positive teen-friendly spaces}
%NOTE: Need a better structure overall hmmm 
%\begin{tcolorbox}[colback=gray!10, colframe=gray!50, width=\textwidth, sharp corners=south, boxrule=0.5mm, title=]
%\textbf{What are design implications for fostering safe and positive online environments?}
%\end{tcolorbox}
 In order to achieve the growth described above, communities must \rv{establish robust frameworks that prioritize }stability and safety. In this section, we identify two key takeaways for fostering such an environment: continuous guidance, empowering teen developers to handle challenges, and collaborative stakeholder involvement. 

\subsubsection{\textbf{Empowering teen developers to handle challenges.}}
\rv{The decentralized nature of developer communities on Roblox creates both opportunities and challenges for teen developers. While these communities foster creativity and peer-to-peer learning, they also leave teens vulnerable to risks such as financial conflicts, copyright issues, and scams, which can be difficult to navigate without guidance. Instead of adopting a top-down governance model that could undermine teen agency and autonomy in managing their spaces, Roblox should focus on empowering teen developers to handle these challenges effectively.}
\rv{Roblox can play a critical role by educating developers about community norms and setting examples of good community policies—not just showcasing successful games but also showcasing communities with successful governance.} Clear rules addressing financial disputes and intellectual property concerns should be established, alongside robust reporting and peer-review mechanisms for identifying and resolving suspicious activities. 

Teen developers should also be encouraged to collaborate in creating solutions \rv{as a primary stakeholder}. For example, \rv{developer communities use }peer review \rv{mechanisms} to identify scams, share knowledge about recognizing common \rv{scam patterns}, and clarify realistic commission expectations to prevent exploitation. Roblox could further support community leaders by providing educational resources on effective moderation strategies and creating incentives for experienced members to mentor newcomers. AI tools could assist by explaining ongoing conversations to newcomers, addressing accessibility challenges, and easing their integration into the community. By focusing on \textit{empowerment} \rv{of teen developers} rather than on control, Roblox can foster resilience-based learning~\cite{park2023towards}, \rv{and community sustainability} as members come in and out. Though it is impossible to eliminate all risks, equipping teen developers with the tools, knowledge, and community support to navigate the challenges of the community can help them grow both creatively and responsibly.

\subsubsection{\textbf{Continuous stakeholder support both offline and online.}}
\rv{Teen developers demonstrate remarkable agency within their communities, which leads to collaborative learning and self-directed socialization. The unique opportunities offered by platforms like Roblox—ranging from exploring diverse games and monetization opportunities to connecting with peers—serve as a gateway for broader offline engagement. For some participants, online experiences inspired decisions to attend relevant schools or participate in offline meetups, while for some participants, their school friends led them to join online communities in the first place. This depicts the organic interplay between the digital and physical worlds~\cite{taylor2009play}. Additionally}, participants emphasized \rv{how} supportive \rv{online} environments often benefit from robust offline support networks, \rv{including teachers, family, and peers, which play a critical role in fostering their growth and helping them navigate challenges. This seamless transition between online and offline support highlights the need for a cohesive, collaborative approach among stakeholders.}

\rv{To nurture teen developers effectively, all stakeholders—parents, educators, peers, platform providers, and developers themselves—must work together to build an ecosystem that bridges online creativity with offline guidance. } To build a more broadly supportive ecosystem, Roblox must position itself as a serious platform for technical development, earning respect from both users and external stakeholders. By integrating Roblox’s educational value in prioritizing \rv{developer} safety, Roblox can strengthen its supportive role. Likewise, parents and schools should recognize \rv{and acknowledge} the significance of Roblox development when students showcase their achievements. Bridging online creativity with offline guidance through this continuous support network is crucial for fostering a positive environment for teen developers.

Effective management of potential issues on Roblox demands active involvement from all stakeholders: Roblox, host platforms for developer communities, developer community moderators, parents, schools, and teen developers. To create an inclusive environment, it is crucial to implement participatory governance models that ensure every voice, especially those of teen \rv{developers}, is heard in shaping community policies~\cite{wang202312}. This collaborative approach may involve regular feedback sessions, open forums, and advisory panels where all stakeholders can contribute to developing community policies and support mechanisms. By fostering these participatory practices, stakeholders can enhance the \rv{community} experience, address emerging issues, and build a more supportive and engaging environment for teen developers.  Such efforts can also reduce the normalization of harmful game design within the Roblox ecosystem, mitigating one of the platform’s most significant risks to players and fostering a safer, more constructive space for all users. 

\subsection{Future Work} %Shorten
Our study reveals promising growth opportunities for teenagers in game development, socializing, and career building in platforms like Roblox but also uncovers several challenges. To address these, future research should aim to include a more diverse sample or utilize large-scale chat data analysis from developer communities. This would provide a broader perspective on the Roblox creator ecosystem and gain insights into underrepresented developers.
While issues like child safety and grooming were minimally reported in our sample, they may be more significant in a larger study. In addition, expanding to explore other teenage communities—such as social media creators or school-based learning platforms—could reveal different motivations and interaction styles due to their unique contexts. Understanding these differences, including how to better support child creators, remains an important area for further investigation.

\section{Conclusion}

For teenagers, creating games is a valuable and engaging method for fostering skills in computational thinking, collaboration, and creative problem-solving. Despite the increased accessibility of game development, many \rv{teen} developers encounter challenges due to the diverse skill sets required. Based on interviews with 18 teen Roblox \rv{developers}, we show how online communities can provide significant benefits, including access to learning resources, opportunities for collaboration, and the growth of transferable skills beyond the gaming platform. However, these communities also present challenges, often stemming from problematic behaviors exhibited by other users within the community, unsupportive guardians, or community environment factors such as school-life balance. These findings highlight the need for more structured guidance and best practices to maximize the positive impacts of online communities and provide a safer space for \rv{teen developers} in their developmental journeys. We hope our work can further spur discussion on the nuances of designing positive online social experiences for \rv{teen developers}. 

\clearpage

\begin{acks}
This research was supported by a grant from the National Research Foundation of Korea (No. RS-2024-00348993). We gratefully acknowledge the Roblox developer community members for their invaluable insights and active participation throughout our research. We also thank @tlllik and @GODPOTATO\_Official, who contributed their creative works shown in Figure \ref{fig:creation}(a) and (c), respectively, and wish to be acknowledged.

\end{acks}

%\bibliographystyle{ACM-Reference-Format}
%\bibliography{reference}


\clearpage
%TC:ignore
\begin{appendix}

\section{Appendix}

\subsection{\rv{Final Themes}}
\label{appendix:codebook}

\begin{figure*}[bh]
    \centering
    \includegraphics[width=1.0\linewidth]{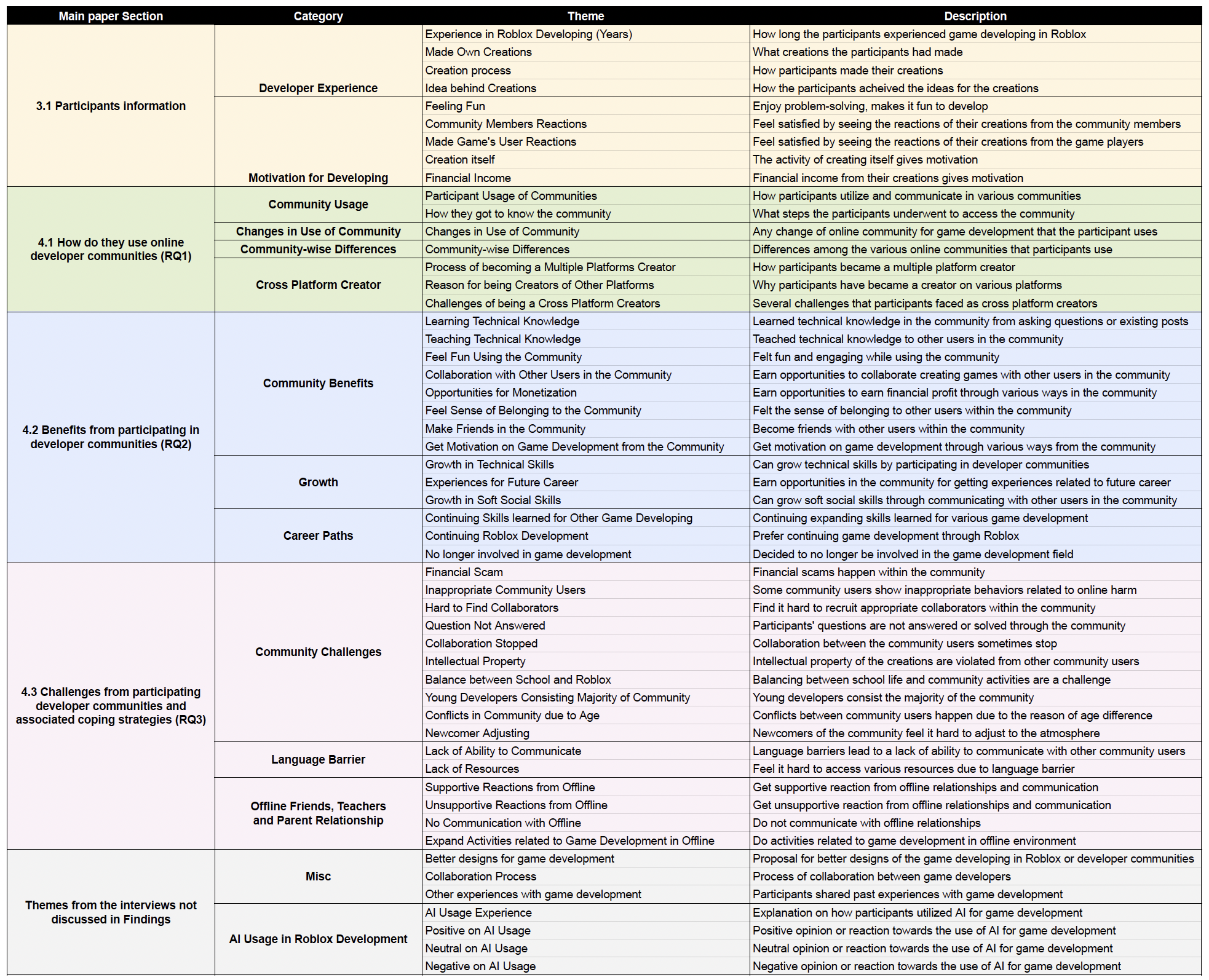}
    \caption{The final themes from \rvtwo{interviews with 18 teen Roblox developers.} \rvtwo{The themes are organized into three major findings areas: (1) Community Use Patterns (RQ1, Community Usage, Changes in Use, Developer Experience), (2) Community Benefits (RQ2, Growth, Community Benefits, Career Paths), (3) Community Challenges (RQ3, Community Challenges, Language Barriers, Offline Relationships). Each category connects to corresponding sections in the Main Paper, including a few themes that emerged from the interviews but weren't central to our findings.}}
    \Description{The themes are organized into three major findings areas: (1) Community Use Patterns (RQ1, Community Usage, Changes in Use, Developer Experience), (2) Community Benefits (RQ2, Growth, Community Benefits, Career Paths), and (3) Community Challenges (RQ3, Community Challenges, Language Barriers, Offline Relationships). Each category connects to corresponding sections in the Main Paper, including a few themes that emerged from the interviews but weren't central to our findings.}
    \label{fig:final_codebook}
\end{figure*}

Figure~\ref{fig:final_codebook} presents the main themes from the interviews. \\

%\clearpage
\subsection{Full Interview Questions}
\label{appendix:questions}
The interview started by explaining this research and asking for consent (and communicating that participants could drop out anytime if they felt uncomfortable). It was semi-structured, beginning with a core set of questions. Questions were modified or added in response to participants' answers, exploring interesting threads more deeply.

\noindent\textbf{Warm-up}
\begin{enumerate}[label=\arabic*., topsep=1pt, noitemsep]
    \item What’s your favorite game in Roblox?
    \begin{enumerate}
        \item What is the genre?
        \item Do you play with friends?
        \item What would you recommend for a newbie?
    \end{enumerate}
    \item Can you show me your memorable Roblox avatar you liked?
    \item Who are your favorite Roblox YouTubers or TikTok influencers?
\end{enumerate}

\noindent\textbf{Becoming a game creator}
\begin{enumerate}[label=\arabic*., topsep=1pt, noitemsep]
    \item What’s the coolest item/game you’ve made in Roblox?
    \begin{enumerate}
        \item Did the idea for the item/game just come into mind?
    \end{enumerate}
    \item Could you tell me more about how you made the item/game?
    \item Have you collaborated with anyone when making the item/game?
    \item How many years did it take to make the item/game?
    \item How many items/games have you made? Can you show me some?
\end{enumerate}

\noindent\textbf{Community-Related}
\begin{enumerate}[label=\arabic*., topsep=1pt, noitemsep]
    \item When you made an item/game, were there times you were stuck?
    \item Where did you go when you’re stuck?
        \begin{enumerate}
            \item Did you seek help online?
            \item How often did you go there?
        \end{enumerate}
    \item Now, could you recall the 1st time you made something in Roblox? Where did you get help then?
        \begin{enumerate}
            \item How did you get to know there?
        \end{enumerate}
    \item Did you have any struggle when you 1st joined \textit{(channel mentioned during the interview)}?
        \begin{enumerate}
            \item Have you ever felt awkward or confused as a newbie?
        \end{enumerate}
    \item (If the communities used have changed) Why do you now go to different communities now?
        \begin{enumerate}
            \item How do you spend your time?
        \end{enumerate}
    \item (If the communities used have not changed) Why do you use the same channel?
        \begin{enumerate}
            \item What do you like about it?
            \item Have you thought about joining others?
            \item How do you spend your time there?
        \end{enumerate}
    \item Do you find it easy to communicate your intention through these channels now?
        \begin{enumerate}
            \item Are you happy with the way the \textit{(channel mentioned during the interview)} is now?
        \end{enumerate}
    
    \item Looking back, for a newbie trying to make an item/a game, where would you recommend?
    \item What would you miss if \textit{(channel mentioned during the interview)} didn’t exist?
    \item Do you feel you’re learning and achieving your goals by being part of \textit{(channel mentioned during the interview)}?
    \item Do you think \textit{(channel mentioned during the interview)} has influenced your work in Roblox?
        \begin{enumerate}
            \item Has it influenced your real life too?
        \end{enumerate}
\end{enumerate}

\noindent\textbf{Safety-Related}
\begin{enumerate}[label=\arabic*., topsep=1pt, noitemsep]
    \item Do you know others’ ages in those \textit{(channel mentioned during the interview)}?
    \item Have you felt someone is older or younger in \textit{(channel mentioned during the interview)}?
        \begin{enumerate}
            \item Why?
            \item Have you ever talked about your age?
        \end{enumerate}
    \item Have you ever felt suspicious of \textit{(channel mentioned during the interview)} as a newbie? Or even now?
    \item How close do you feel to the other members of the channel?
    \item Do you think there should be a separate channel just for kids?
\end{enumerate}

\noindent\textbf{Offline-Related}
\paragraph{Thank you so much. I would like to move on to the last topic.}
\begin{enumerate}[label=\arabic*., topsep=1pt, noitemsep]
    \item Have you shown what you made in Roblox to your parents or offline friends? Why or why not?
\end{enumerate}

\noindent\textbf{Concluding thoughts}
\begin{enumerate}[label=\arabic*., topsep=1pt, noitemsep]
    \item What are some things Roblox developer communities could do better?
    \item Do you plan to continue developing in Roblox?
    \item Is there anything you want to share with us?
\end{enumerate}

\end{appendix}
%TC:endignore

\end{document}